\documentclass{WileyMSP-template}
\usepackage[version=3]{mhchem} 
\usepackage{subcaption}
\usepackage{graphicx}
\usepackage{amsmath}
\usepackage{amssymb}
\usepackage{bm}
\usepackage{textgreek} 
\usepackage{chemformula}
\usepackage[nopatch]{microtype}
\usepackage{booktabs}
\usepackage{float}
\usepackage[english]{babel}
\usepackage[
    colorlinks=true,
    linkcolor=black,
    urlcolor=black,
    citecolor=black,
    bookmarks=false   
]{hyperref}

\begin{document}


\title{\textbf{Photoluminescence Mapping of Mobile and Fixed Defects in Halide Perovskite Films}}

\maketitle


\author{Sarah C. Gillespie\textsuperscript{1,}\textsuperscript{2}},
\author{J\'{e}r\^{o}me Gautier}\textsuperscript{1},
\author{Linde M. van de Ven}\textsuperscript{1},
\author{Agustin O. Alvarez}\textsuperscript{1},
\author{Bruno Ehrler}\textsuperscript{1},
\author{L.J. Geerligs}\textsuperscript{2},
\author{Veronique S. Gevaerts}\textsuperscript{2},
\author{Gianluca Coletti}\textsuperscript{2,}\textsuperscript{3}, and
\author{Erik C. Garnett*}\textsuperscript{1,}\textsuperscript{4}

\begin{affiliations}
\textsuperscript{1} LMPV-Sustainable Energy Materials Department, AMOLF Institute, Science Park 104, Amsterdam, 1098XG, The Netherlands \\
\textsuperscript{2} TNO Department Solar Energy, Westerduinweg 3, Petten, 1755LE, The Netherlands\\
\textsuperscript{3} School of Photovoltaic and Renewable Energy Engineering, University of New South Wales, Sydney, New South Wales 2052, Australia\\ 
\textsuperscript{4} Institute of Physics, University of Amsterdam, Science Park 904, Amsterdam, 1098XH The Netherlands\\ 
{*} Corresponding author. Email: e.garnett@amolf.nl
\end{affiliations}



\begin{abstract}
Metal halide perovskites exhibit coupled electronic and ionic properties that determine their photovoltaic performance and operational stability. Understanding and quantifying ionic transport are therefore essential for advancing perovskite optoelectronics. Conventional electrical methods such as impedance spectroscopy require fully integrated devices, and their interpretation is often complicated by interfacial and contact effects, limiting the ability to isolate intrinsic ionic behavior. Here, a localized adaptation of intensity-modulated photoluminescence spectroscopy (IMPLS) is utilized to optically probe lateral ionic transport in perovskite films. The frequency-dependent photoluminescence response is measured under controlled carrier injection levels and correlated with the photoluminescence quantum yield (PLQY). The proposed diffusion model indicates that mobile ionic defects laterally migrate from high light intensity regions, giving rise to characteristic photoluminescence modulations. Ionic diffusion coefficients extracted from IMPLS agree well with literature values obtained from electrical measurements. Importantly, IMPLS mapping separates mobile and immobile defect contributions through a defect contrast coefficient (DCC), which quantifies the normalized difference between the area-averaged photoluminescence intensity and phase data. This work ultimately demonstrates that localized IMPLS provides a contact-free means to extract lateral ion diffusion
coefficients while spatially distinguishing defect types across the sample.
\end{abstract}


\section{Introduction}
Metal halide perovskites are among the most promising materials for next-generation semiconductor applications owing to their exceptional properties such as high absorption coefficients, long carrier diffusion lengths and long carrier lifetimes\cite{perovskiteforfuture,abscoeff,umdiff,us1,topo}. These properties have made them particularly attractive for high-efficiency photovoltaics (PV)\cite{tables,progresspsc}. Beyond PV, perovskites have potential applications in a wide range of technologies, including transistors, memristors, photo- and X-ray detectors, self-tracking solar concentrators, and more\cite{transistor,memristor1,memristor2,memristor3,jeroen,memory,memlumor,flexphotodet,sciencephotodet,Fang2015,xrayperov,juliasolarconc,beyond}. Many of these emerging applications exploit the coupled electronic–ionic properties inherent to the perovskite material itself. For example, halide segregation drives the self-tracking behavior in solar concentrators, while the operating principle of perovskite memristors relies on ionic migration and the associated memory effect, enabling information retention over extended periods and the formation of multiple resistive states within a device\cite{phaseseg,juliasolarconc,jeroen,ionbenefits,explanationpermem}. \\ Among the available characterization tools, photoluminescence (PL) spectroscopy stands out as one of the most widely used for probing electronic charge-carrier processes in the perovskite community\cite{photokir}. By tuning the excitation conditions such as energy, intensity, and repetition rate, PL measurements can be analyzed to determine key electronic properties, including carrier lifetimes, surface recombination velocities and carrier diffusion lengths\cite{herztraps,gingertrpl,underkirchartz,siliconinspired,Cho2022}. PL time series have also been applied to study ionic phenomena such as halide segregation and mobile ion redistribution -- however, the resulting information from these measurements has been largely limited to qualitative trends\cite{superoxide,controlling,photobright,phaseseg}. Quantitative methods for determining ionic parameters, such as mobile ion densities and diffusion coefficients, are still predominantly in the domain of electrical measurements; techniques including transient ion drift and impedance spectroscopy (IS) are commonly applied to extract these ionic values\cite{futscher2020,moritz1,lucie2021,Peng2018}. Despite this, all electrical approaches face inherent limitations: measurements are restricted to fully operational devices, while high recombination at the transport layers and interfacial ionic reactions tend to complicate the interpretation of results\cite{moritz2,contactsis}.\\

Recently, we introduced an adaptation of modulated PL spectroscopy -- called intensity-modulated photoluminescence spectroscopy (IMPLS) -- specifically designed to resolve ion dynamics purely optically in perovskite films and devices\cite{mplsilicon2006,IMPLS}. Unlike time-domain PL measurements, where overlapping processes complicate interpretation, IMPLS separates contributions with distinct relaxation times by probing the PL response in the frequency domain. This frequency-dependent PLQY response raises a key question: how can frequency modulations on the timescale of seconds influence the PLQY, a parameter typically governed by nanosecond-to-microsecond charge-carrier recombination dynamics? We address this question in this work by correlating the frequency-dependent PL response with absolute PLQY measurements under localized excitation conditions. Specifically, we measure IMPLS on a halide perovskite thin film both as a function of frequency ($f$) and of excess minority carrier density ($\Delta n$), and we directly compare the phase component of the IMPLS measurements to the PLQY across the same $\Delta n$ and $f$ range. We then spatially resolve these correlations across a 625 \textmu m\textsuperscript{2} area on the sample. From this analysis, we propose a model of the coupled electronic–ionic processes, in which the IMPLS signatures originate from lateral mobile ion diffusion between the laser-illuminated region and the surrounding background. The analysis indicates that, under localized experimental conditions, IMPLS can be applied as a means to determine mobile ion diffusion coefficients ($D_{\textrm{ion}}$) entirely without the requirement of electrical contacts. In the case where the total defect density in the halide perovskite is dominated by mobile ionic defects, the lateral diffusion coefficient can be correlated to the change in the PLQY\cite{fenning}. Thus, analyzing differences between IMPLS and PLQY maps provides a route to spatially resolve regions where non-radiative recombination is dominated by either mobile or fixed defects.

\section{Time-Dependent PLQY in Halide Perovskites}
\begin{figure}[!ht]
\centering
\includegraphics[width=0.7\linewidth]{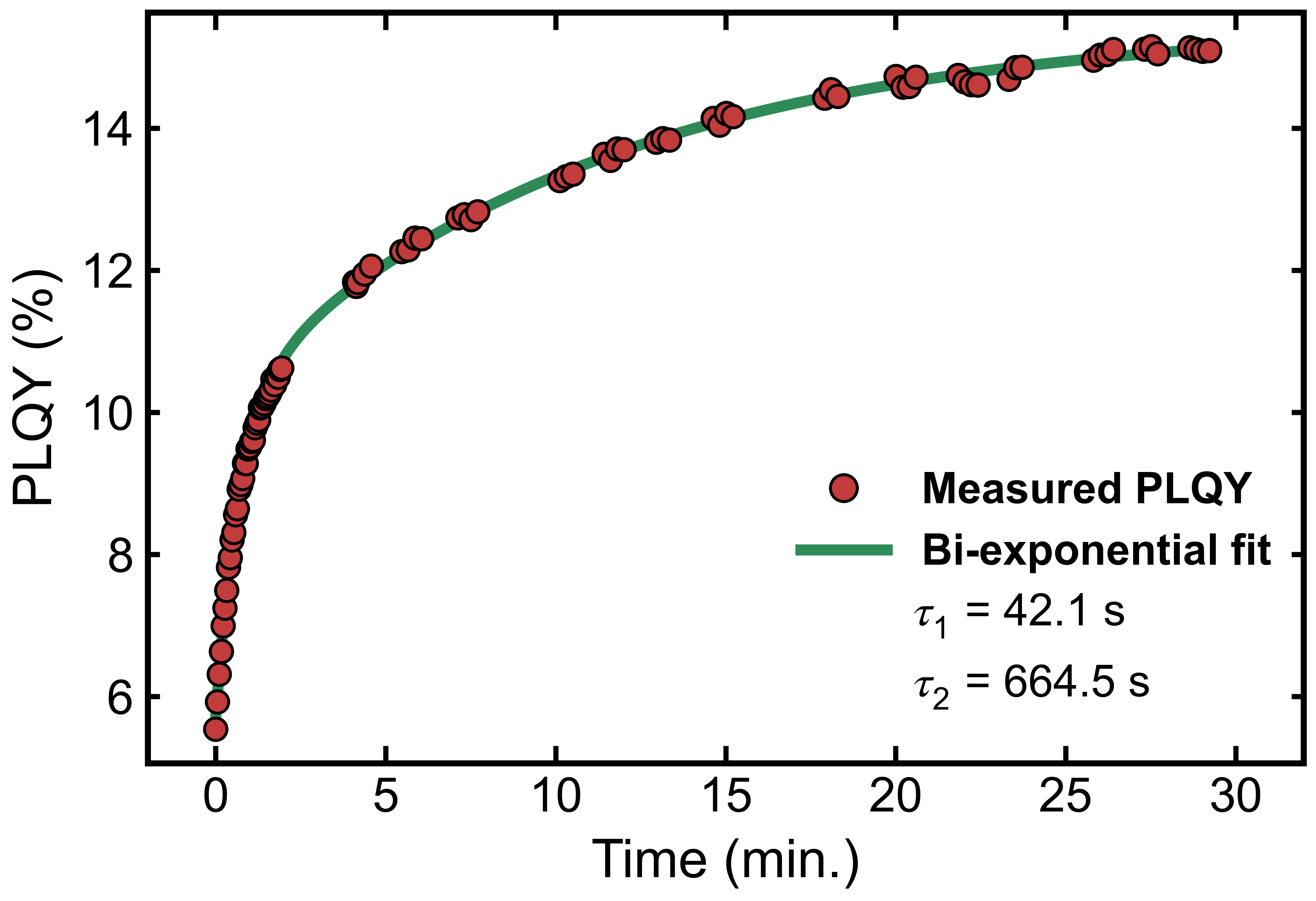}
\caption{PLQY time series measured on an encapsulated triple-cation, mixed-halide perovskite film. The empty regions between the red data points are when the sample's absorptance was measured (see the SI for measurement details). The green curve represents the bi-exponential rise fit to the data, $\textrm{PLQY} = \textrm{PLQY\textsubscript{saturation}} - A\exp(-t/\textrm{\texttau \textsubscript{1}}) - B\exp(-t/\textrm{\texttau \textsubscript{2}})$. The PLQY saturation value from the fit is 15.5\% and the extracted time constants are \texttau \textsubscript{1} = 42.1 seconds and \texttau \textsubscript{2} = 11.08 minutes.}
\label{fig1}
\end{figure}

Before discussing the IMPLS results, we first consider the PLQY variations observed in halide perovskites under time-domain measurements, and elucidate how these effects manifest in the frequency domain. Assuming no processes alter the semiconductor's electronic properties under continuous illumination, the PLQY is defined as the ratio of the radiative recombination rate, $R_{\textrm{rad}}$, to the total recombination rate under steady-state conditions:
\begin{equation}\label{simplePLQY}
    \textrm{PLQY} = \frac{R_{\textrm{rad}}}{R_{\textrm{rad}} + R_{\textrm{trap}} + R_{\textrm{Auger}}}
\end{equation}
Here, $R_{\textrm{trap}}$ and $R_{\textrm{Auger}}$ are the trap-assisted and Auger non-radiative recombination rates, respectively. In reality, the PLQY of halide perovskites can vary in time even under moderate illumination conditions. In \textbf{Figure \ref{fig1}}, we plot the measured PLQY time series of a \ce{Cs_{0.07}(FA_{0.8}MA_{0.2})_{0.93}Pb(I_{0.8}Br_{0.2})3} thin film encapsulated in \ce{SiO2}. Sample fabrication and measurement details are listed in the Supporting Information (SI). We observe a rapid PLQY rise followed by a gradual plateau as the measurement time approaches 30 minutes. While this example exhibits a relatively straightforward bi-exponential rise, many studies have shown that the PL of perovskite films can alternate between photobrightening and photodarkening, depending on timescale and measurement conditions\cite{marialoi1,platmosphere,shiningdavies,realtime}. Notable ionic processes that drive photobrightening or photodarkening include Frenkel pair generation and recombination, formation of complex species under specific illumination conditions and ionic trap migration out of the illuminated region\cite{photobright,frenkel,controlling,superoxide}. To account for these changes under steady-state or quasi-steady-state excitation, we introduce an additional recombination term, $R_{\textrm{slow}}(t)$, in the PLQY definition. This time-varying non-radiative term -- which can be positive or negative -- encompasses changes in trap-state density, recombination coefficients, or other parameters that modify the total recombination rate over time. The time-dependent PLQY can thus be expressed as:
\begin{equation}\label{eq:timePLQY}
    \textrm{PLQY}(t) = \frac{R_{\textrm{rad}}}{R_{\textrm{rad}} + R_{\textrm{trap}} + R_{\textrm{Auger}}+R_{\textrm{slow}}(t)}
\end{equation}
Provided that the slow process affects only the minority-carrier lifetime, \textbf{Equation \ref{eq:timePLQY}} can instead be written with a time-dependent trap-assisted recombination rate, $R_{\textrm{trap}}(t)$. As illustrated in \textbf{Figure \ref{fig1}}, the ratio of radiative to non-radiative recombination increases, leading to an enhanced PLQY. This enhancement may arise from, for example, ionic defects diffusing out of the illuminated region or from vacancy–interstitial recombination, resulting in a temporal decay in the overall non-radiative recombination rate\cite{photobright,controlling}.\\
In the corresponding frequency-domain analysis, the PLQY can vary depending on the applied excitation frequency. If the optical modulation frequency is much faster than the characteristic lifetime of the ionic process, $1/2\pi f \ll \textrm{\texttau \textsubscript{char}}$, then the process cannot follow the modulation and therefore does not introduce any additional change to the PL phase shift or amplitude -- the PL signal simply follows the imposed input modulation, without any additional frequency-dependent modification from the ionic process. This effect is analogous to the ``ion-freeze'' concept for perovskite solar cell $JV$ sweeps, which occurs when the $JV$ curve is measured at scan rates faster than the ionic response time\cite{jarla2024}. In that case, the mobile ion contribution is suppressed at high $JV$ scan rates, whereas at slower sweeps, their contribution becomes increasingly dominant in the $JV$ response. Similarly, at high modulation frequencies, ionic effects are frozen, resulting in an unchanged PLQY. Once the modulation frequency is sufficiently low to become comparable to the characteristic timescale of an ionic process, its influence appears through a measurable phase shift and amplitude change in the IMPLS measurements, corresponding to a changing $R_{\textrm{slow}}(t)$ in \textbf{Equation \ref{eq:timePLQY}}, thus a change in the PLQY. When the modulation frequency is well below the probed process, the system behaves as if under quasi steady-state (DC) conditions, since the process can fully equilibrate within each cycle. This implies that the low frequency IMPLS response and PLQY changes on the timescale of seconds are both driven by the same ionic process. In this work, we test this hypothesis by directly correlating our measured IMPLS responses with absolute PLQY measurements. 

\section{IMPLS Results}
\subsection{Experimental Configuration and Measurement Analysis}
\begin{figure}[!ht]
\centering
\includegraphics[width=0.9\linewidth]{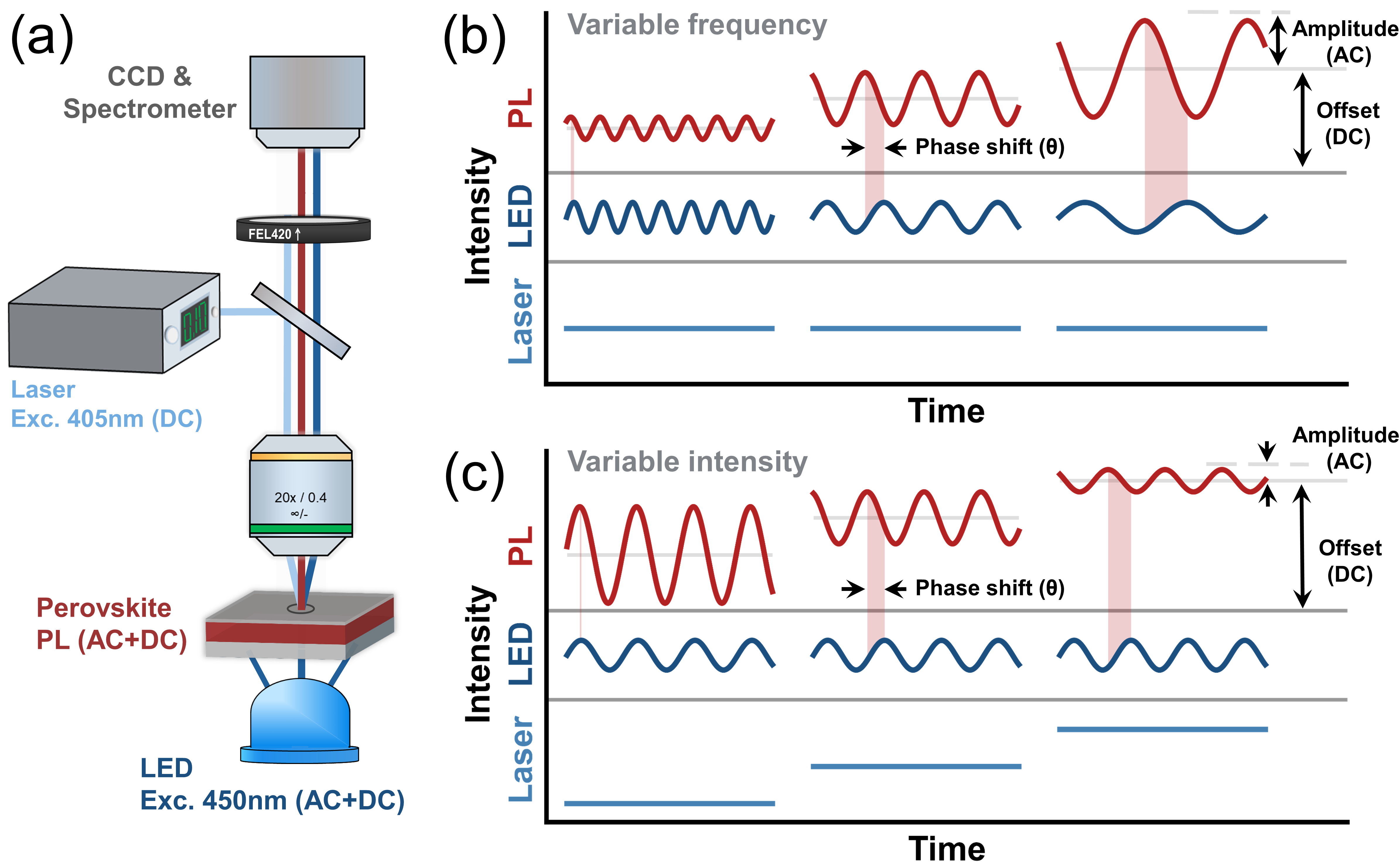}
\caption{(a) Schematic of the experimental setup. Laser excitation was applied to the top of the sample via a dichroic mirror and its reflection was blocked using a 420 nm long-pass filter. The LED provided an offset intensity of 38.1 mW/cm\textsuperscript{2}, corresponding to a photon flux of \textPhi \textsubscript{DC,LED} = 8.63$\times$10\textsuperscript{16} cm\textsuperscript{-2}/s. The LED amplitude was set to 15.7 mW/cm\textsuperscript{2} (\textPhi \textsubscript{AC,LED} = 3.56$\times$10\textsuperscript{16} cm\textsuperscript{-2}/s). For experiment (i) -- the variable frequency experiment -- the laser intensity was fixed at 33.104 W/cm\textsuperscript{2} (\textPhi \textsubscript{laser} = 6.75$\times$10\textsuperscript{19} cm\textsuperscript{-2}/s). For experiment (ii) -- the variable intensity experiment -- the LED frequency was fixed at 50 mHz. (b) Schematic of experiment (i) for three representative modulation frequencies at a fixed offset intensity. As the modulation frequency decreases (dark blue LED curve), the PL AC amplitude and DC offset increase (red). The relative phase shift (exemplified by the shaded regions from the PL peak to the subsequent LED peak) also increases with decreasing frequency. (c) Schematic of experiment (ii) for three representative laser intensities at a fixed modulation frequency. Increasing laser intensity (light blue laser line) similarly increases the PL DC offset and phase shift. The PL AC amplitude decreases for increasing intensity. The amplitudes, offsets and phase shifts are exaggerated for clarity but follow the same trends observed experimentally in this work.}
\label{fig2}
\end{figure}

The experimental setup used for the IMPLS measurements is schematically illustrated in \textbf{Figure \ref{fig2}a}. A perovskite sample, identical in composition to that shown in \textbf{Figure \ref{fig1}}, was placed above a 450 nm light-emitting diode (LED), which globally illuminated the sample. The LED provided an excitation density consisting of a sinusoidal (AC) modulation, $\widetilde{\phi}_{\textrm{exc,LED}}$, superimposed on a constant offset (DC) excitation, $\bar{\phi}_\textrm{exc,LED}$. The instantaneous LED excitation density at time $t$ can be expressed as:
\begin{equation}
    \phi_\textrm{exc,LED}(t) = \bar{\phi}_\textrm{exc,LED} + |\widetilde{\phi}_\textrm{exc,LED}| \sin(2\pi f t )
\end{equation}
where $|\widetilde{\phi}_\textrm{exc,LED}|$ represents the sinusoidal modulation amplitude. \\ In parallel, a 405 nm laser provided additional high-intensity, localized excitation, ${\phi}_\textrm{exc,laser}$, focused onto a spot area of 78.5 µm\textsuperscript{2}. The total excitation density at the laser spot was therefore given by the sum of the LED and laser contributions:
\begin{equation}
{\phi}_\textrm{exc}(t) = {\phi}_\textrm{exc,laser} + {\phi}_\textrm{exc,LED}(t)
\end{equation}
As illustrated in \textbf{Figure \ref{fig2}a}, the PL of the sample was collected through a 20$\times$ objective and recorded using a charge-coupled device (CCD) coupled to a spectrometer. The spectra were converted to the energy scale, and both the LED and PL signals were integrated and tracked in time\cite{jacobian}. The PL signal can thus be described as:
\begin{equation}\label{eq:PLSin}
    \phi_\textrm{PL}(t) = \bar{\phi}_\textrm{PL} + |\widetilde{\phi}_\textrm{PL}| \sin(2\pi f t + \theta)
\end{equation}
where $\bar{\phi}_\textrm{PL} $ is the offset PL intensity (the DC component), $|\widetilde{\phi}_\textrm{PL}|$ is the amplitude of the PL modulation (the AC component), and $\theta$ denotes the relative phase shift between the modulated illumination input and the PL output. By fitting sinusoidal functions to the integrated LED and PL spectra, all three parameters were determined. \\ An important distinction between the IMPLS analytical approach used here and conventional frequency-domain electrical analyses (such as IS) is that in our case, the DC component of the PL response is not constrained to remain constant with modulation frequency. In electrical IS, maintaining a constant DC response is essential to ensure that the system remains in a quasi-steady state during small-signal perturbation\cite{hauffemerging,ISnoise}. In contrast, our measurements intentionally capture both the DC and AC components of the PL as a function of frequency and excitation density, since changes in $\bar{\phi}_\textrm{PL}$ also contain information about carrier population dynamics under modulated excitation. To ensure that observed DC variations are intrinsic and not due to sample degradation or measurement history, each frequency point was measured on a fresh, spatially distinct region of the sample, and the measurement sequence was fully randomized\cite{julia,memlumor}. This approach rules out cumulative effects or measurement artifacts. Further details of this analysis are discussed in the SI. \\
With the measurement procedure established, we conducted two key IMPLS experiments: (i) in the first experiment, we varied the LED modulation frequency while holding the offset illumination intensity constant, and recorded the IMPLS response as a function of $f$ (\textbf{Figure \ref{fig2}b}); (ii) in the second experiment, we varied the excess minority carrier density by sweeping the laser intensity at a fixed $f$, and measured the IMPLS response as a function of $\Delta n$ (\textbf{Figure \ref{fig2}c}). Although the scales in \textbf{Figure \ref{fig2}b} and \textbf{Figure \ref{fig2}c} are exaggerated for visualization, the changes in PL phase shift, amplitude, and offset follow the trends that were experimentally obtained in this work.

\subsection{IMPLS Response as a Function of Modulation Frequency}
\begin{figure}[!ht]
\centering
\includegraphics[width=0.9\linewidth]{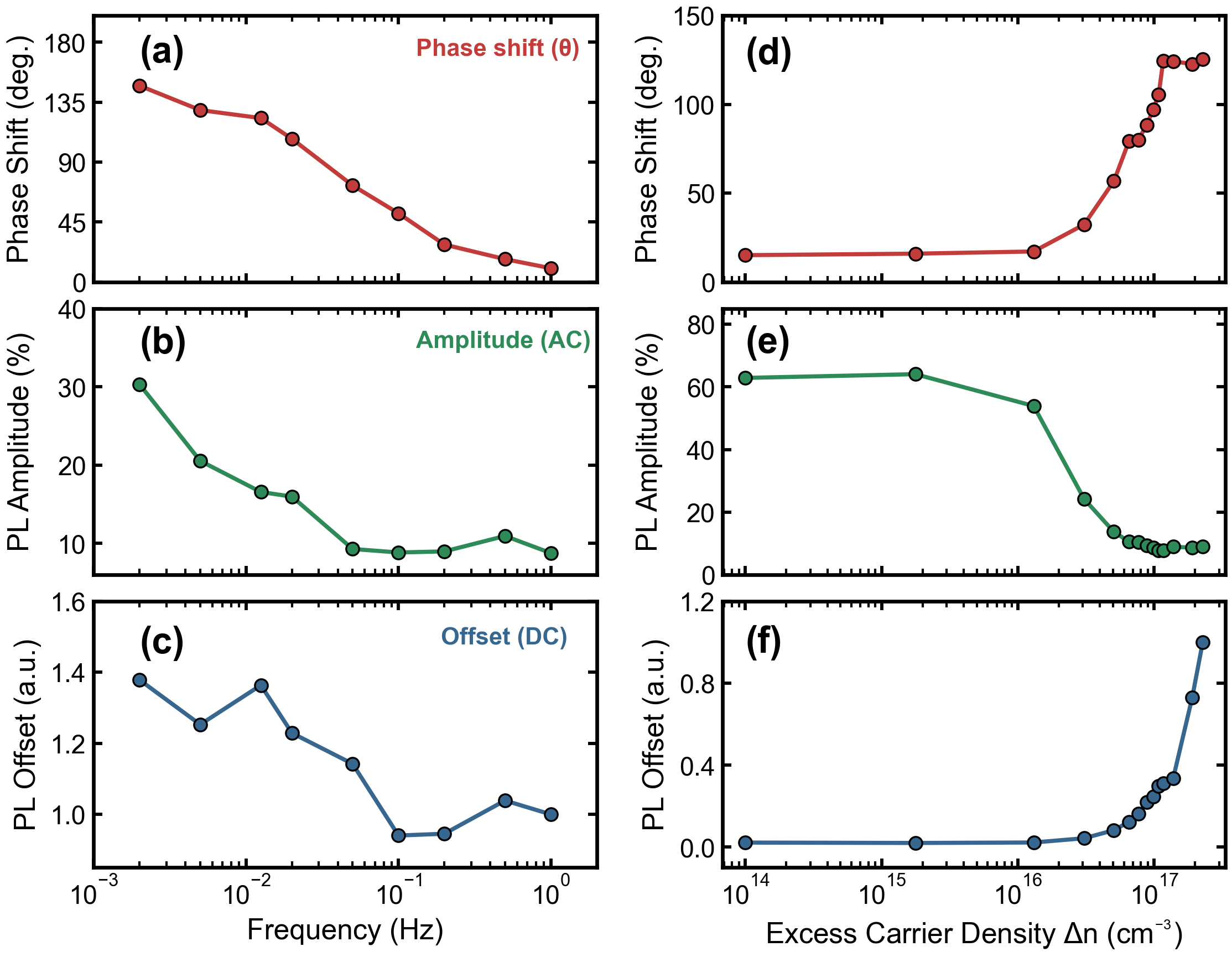}
\caption{Panels (a) - (c) show results from experiment (i), corresponding to \textbf{Figure \ref{fig2}b}. Panels (d) - (f) show results from experiment (ii), corresponding to \textbf{Figure \ref{fig2}c}. (a) Bode plot of the PL phase shift as a function of LED modulation frequency. (b) PL amplitude and (c) PL offset Bode plots over the same frequency range. The PL offset was normalized to the PL offset signal collected at $f =$ 1 Hz. (d) Bode plot of the PL phase shift as a function of the excess minority carrier density at a fixed modulation frequency of $f$ = 50 mHz. The carrier density was calculated from the combined photon flux of the laser and LED. (e) Corresponding PL amplitude and (f) PL offset over the same $\Delta n$ range, with the PL offset normalized to the signal collected at maximum $\Delta n$.}
\label{fig3}
\end{figure}

Focusing first on experiment (i), we measured the IMPLS response across modulation frequencies between 2 mHz and 1 Hz. \textbf{Figure \ref{fig3}a} presents the Bode plot of the phase shift across the frequency window; the positive phase shift indicates that the PL amplitude consistently led the LED amplitude across the entire range\cite{hauffis,tutorial}. \textbf{Figure \ref{fig3}b} shows the corresponding PL AC amplitude, and \textbf{Figure \ref{fig3}c} the PL DC offset. The associated uncertainty of the phase, arising from both the fitting procedure and sample inhomogeneity, ranged approximately 1\% and 10\%, depending on the applied frequency. This uncertainty was calculated based on repeated measurements across different sample points (\textbf{Figure \ref{figSIhistogram}}; see the SI for details).\\
Notably, both the PL amplitude and offset increased once $f \lesssim $ 100 mHz. Since the incident photon flux remained constant across all frequencies, this implies that a slow process with an onset time of \texttau\textsubscript{onset} = $1/2\pi f  = $ 1.6 s locally improved the sample's PLQY. The increase in the PL DC offset reflects the time-averaged enhancement of the PLQY, whereas the enhancement in the PL AC amplitude quantifies the degree to which the process modulates the PLQY during the frequency sweep. Even at the lowest frequency measured (2 mHz, corresponding to \texttau\textsubscript{meas. limit} = 79.6 s), the sample's PLQY continued to increase. We further discuss this trend and the potential underlying mechanisms later in this work.                                                                                                                                       

\subsection{IMPLS Response as a Function of Excess Carrier Density}
Turning to experiment (ii), the modulation frequency was fixed at $f = 50$ mHz, and $\Delta n$ was varied by adjusting the laser intensity. The resulting incident photon flux (\textPhi) ranged from 7.7$\times$10\textsuperscript{16} cm\textsuperscript{-2}/s (laser off) to 5.6$\times$10\textsuperscript{20} cm\textsuperscript{-2}/s. {\textPhi} was converted to the carrier generation rate $G$ using $G = A\Phi/W_{\textrm{pvk}}$, where the absorptance was $A = 72\pm2$\%, and the film thickness $W_{\textrm{pvk}} =$ 560 nm. The excess carrier density was then obtained as $\Delta n = G\tau_\mathrm{eff}(\Delta n)$, assuming rapid spatial homogenization of the photogenerated excess carriers. This assumption is justified by the relatively high electronic mobilities ($\gtrsim 1$ cm\textsuperscript{2}V\textsuperscript{-1}s\textsuperscript{-1}) and long carrier lifetimes ($\gtrsim 100$ ns) for perovskite films\cite{underkirchartz,herzmobility,Lim2022}. The trap-assisted lifetime $\tau_\mathrm{trap}$ was measured through time-resolved photoluminescence (TRPL) spectroscopy and applying the differential lifetime analysis method\cite{underkirchartz,siliconinspired}. The analysis yielded $\tau_\mathrm{trap} = 100.5$ ns. Together with radiative and Auger recombination coefficients (determined from intensity-dependent PLQY, discussed in the following section), $\tau_\mathrm{eff}(\Delta n)$ was evaluated. \\ Although $f = 50$ mHz was chosen here as a case study, experiment (ii) was repeated across several additional frequencies between $f = 10$ mHz – 1 Hz; these additional results are shown for completeness in \textbf{Figure \ref{figSI5}}. \\
As illustrated in \textbf{Figure \ref{fig3}d}, the IMPLS phase remained constant up to $\Delta n$\textsubscript{threshold} $\approx$10\textsuperscript{16} cm\textsuperscript{-3}, where $\theta \approx$16$^{\circ}$. Above this threshold, the phase increased with $\Delta n$ before saturating at $\Delta n_\mathrm{saturation} \approx 10^{17}$ cm\textsuperscript{-3}. The PL amplitude (\textbf{Figure \ref{fig3}e}) exhibited similar threshold and saturation points, but decreased in magnitude with increasing $\Delta n$.\\ Interestingly, the phase trend is similar to the expected PLQY trend as it transitions from low injection levels (LIL) to high injection levels (HIL), provided that the doping density of the film is on the order of $10^{16} - 10^{17}$ cm\textsuperscript{-3}. Correspondingly, the unchanged phase and amplitude below $\Delta n$\textsubscript{threshold} would be linked to LIL, while $\Delta n$\textsubscript{saturation} would be linked to where Auger recombination begins to dominate over radiative recombination in the PLQY trend.

\section{Phase Correlation with PLQY}
\begin{figure}[!ht]
\centering
\includegraphics[width=0.9\linewidth]{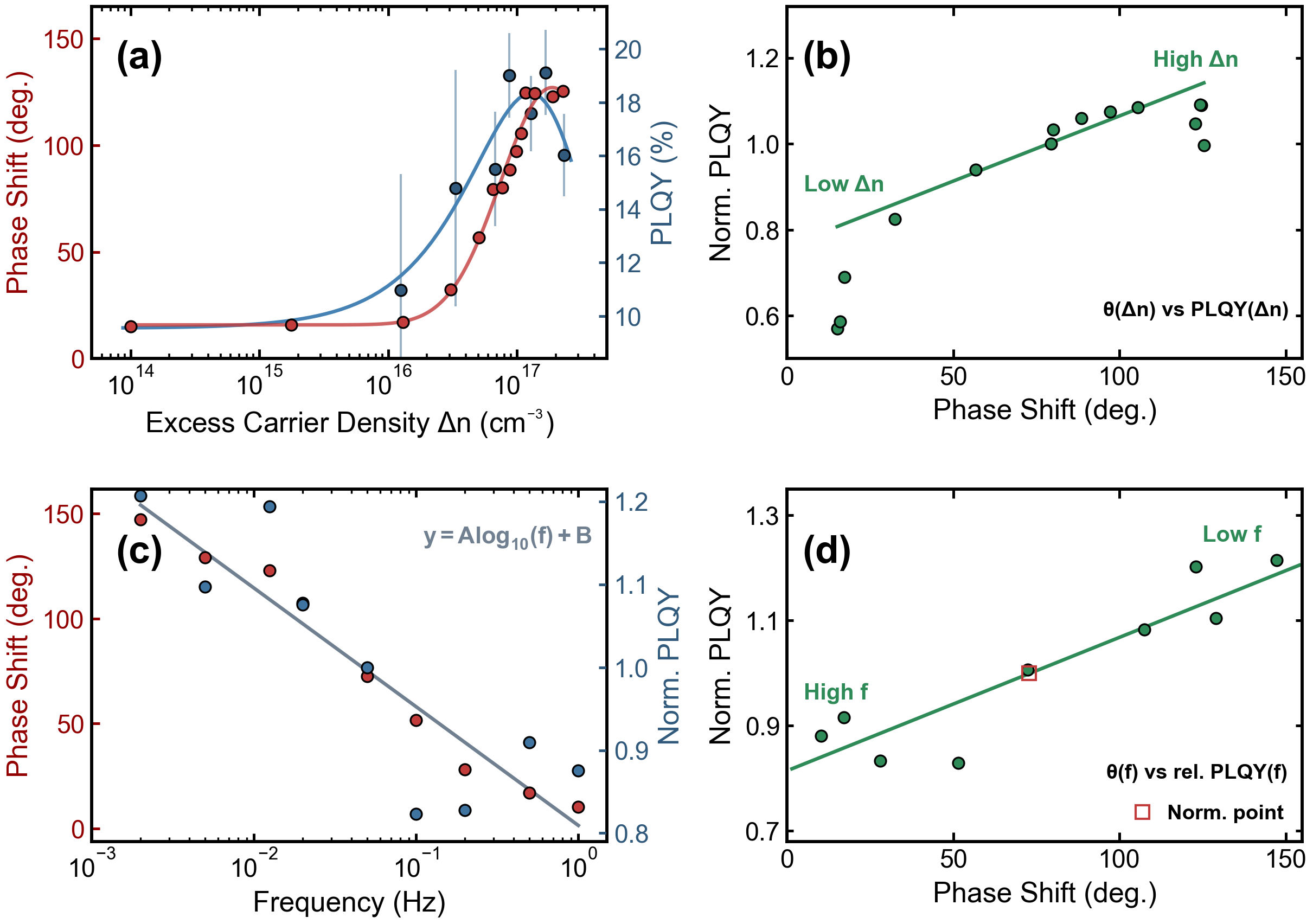}
\caption{(a) PL phase shift data (red markers) as a function of excess minority carrier density (left axis). Averaged PLQY values (25 measurements per point), with standard deviation error bars are shown with the blue markers (right axis). The  blue curve represents the PLQY fit using \textbf{Equation \ref{eq:PLQY}} and red curve represents the phase fit using \textbf{Equation \ref{eq:logspaceG}}. (b) Re-visualizing the data from panel (a) directly as the normalized PLQY versus $\theta$. A linear fit (green line) was applied to the data to visualize the correlated trend. Outliers due to the difference in measurement setups at the high and low $\Delta n$ limits were omitted from the fit. (c) PL intensity phase shift (red, left axis) and corresponding relative PLQY (blue, right axis) as functions of modulation frequency. Both datasets were fitted using a log-linear model, $y(f) = A\log_{10}(f) +B$. (d) Re-visualizing the data from panel (b) directly as the normalized PLQY versus $\theta$. Both datasets in panels (b) and (d) are normalized to their common point (red square) at $f$ = 50 mHz, $\Delta n$ = 6.6$\times10^{16}$ cm\textsuperscript{-3}.}
\label{fig4}
\end{figure}

To quantify this correlation, we measured the absolute PLQY across the same $\Delta n$ range and on the same perovskite sample as that in \textbf{Figure \ref{fig3}}. Localized PLQY maps (25 \textmu m\textsuperscript{2}) were collected as a function of $\Delta n$ using a focused 660 nm laser in a custom-built integrating sphere microscopy setup (see the SI for measurement details). In \textbf{Figure \ref{fig4}a}, we overlay the measured PLQY($\Delta n$) with $\theta$($\Delta n$). The scaling between $\theta$ (primary y-axis, in red) and PLQY (secondary y-axis, in blue) is arbitrary, but supports the correlation hypothesis. Using \texttau\textsubscript{trap} = 100.5 ns, we fit the experimental PLQY data to the analytical model:
\begin{equation}\label{eq:PLQY}
    \textrm{PLQY}(\Delta n) = \frac{k_\textrm{rad}\Delta n(\Delta n+p_0)}{\Delta n/\textrm{\texttau}_\textrm{trap}+k_\textrm{rad}\Delta n(\Delta n+p_0)+C_\textrm{Auger}\Delta n^2(\Delta n+p_0)}
\end{equation}
from which, the radiative and Auger recombination coefficients, and the doping density were determined: \textit{k}\textsubscript{rad} $=( 2.2 \pm 0.97)\times10$\textsuperscript{-11} cm\textsuperscript{3}/s, \textit{C}\textsubscript{Auger} $= (3.2 \pm 2.5)\times10$\textsuperscript{-28} cm\textsuperscript{6}/s and \textit{p\(_{0}\)} $= (4.8 \pm 3)\times10$\textsuperscript{16} cm\textsuperscript{-3}\cite{photokir}. While \textit{C}\textsubscript{Auger} shows some deviation -- most likely due to the limited data points collected in the Auger regime -- \textit{k}\textsubscript{rad} and \textit{p\(_{0}\)} are consistent with our previously reported values for a similar perovskite composition\cite{siliconinspired}. The PLQY fit is shown by the blue curve in \textbf{Figure \ref{fig4}a}.\\ The phase data across the $\Delta n$ range was then fit to a logarithmically-scaled Gaussian function: 
\begin{equation}\label{eq:logspaceG}
    \theta(\Delta n) = A \exp\Big(\frac{-(\log_{10}(\Delta n)-\mu_{\log})^2}{2\sigma_{\log}^2}\Big) +B
\end{equation}
where $\mu_{\log}$ and $\sigma_{\log}$ represent the mean and standard distribution logarithmic space. This empirical fit is shown by the red curve in \textbf{Figure \ref{fig4}a}, with the fit coefficients and their uncertainties listed \textbf{Table \ref{tab:gaussian_params}}. Compared to the PLQY model, the phase shift transitions to its HIL-like regime at a higher $\Delta n$; $\theta$ reaches its maximum at $\Delta n$ = 8.06$\times$10\textsuperscript{17} cm\textsuperscript{-3}, compared to the PLQY peak at $\Delta n$ = 1.23$\times$10\textsuperscript{17} cm\textsuperscript{-3}. While this shift could reflect different processes separately influencing the phase and PLQY, we postulate that it is more likely a consequence of the different setups used for the two measurements. In particular, the beam diameter of the IMPLS configuration was 6.25 times larger than the beam diameter in the PLQY system. As we will show in the next section, the laser beam size significantly impacts the overall IMPLS response. \\
The relationship between the measured PLQY and the phase shift is shown in \textbf{Figure \ref{fig4}b}. We anticipate that under the same measurement conditions, the PLQY scales linearly with $\theta$; this is illustrated by the fit in \textbf{Figure \ref{fig4}b}. The deviations from linearity at the high and low $\Delta n$ limits are a consequence of the $\Delta n$ offset between the curves.\\ 
Relating these observations to the frequency-dependent IMPLS response, if we consider $\theta$ to be directly correlated to the PLQY through ionic processes, then the increasing phase shift measured for decreasing frequency (\textbf{Figure \ref{fig3}a}) should also correspond to a rise in the PLQY. While we do not measure the absolute PLQY as a function of modulating frequency, we can quantify the frequency-dependent relative PLQY as the PL offset intensity in the IMPLS frequency-sweep measurement. Overlaying the relative PLQY with the phase shift show that they indeed both exhibit the same logarithmic dependence on $f$ (\textbf{Figure \ref{fig4}c}), meaning a direct linear correlation to each other. The correlation shown in \textbf{Figure \ref{fig4}d} supports our claim that the PLQY and $\theta$ are linearly correlated also as a function of $f$. Both curves in \textbf{Figure \ref{fig4}b} and \textbf{\ref{fig4}d} are normalized to their common IMPLS point ($f$ = 50 mHz, $\Delta n$ = 6.6$\times10$\textsuperscript{16} cm\textsuperscript{-3}, PLQY = 16.78\%). The overall correlation suggests that IMPLS phase data may be considered a proxy for the PLQY under specific measurement conditions.

 \clearpage

\section{Impact of Beam Size and IMPLS Spatial Maps}\label{sec:maps}
\begin{figure}[!ht]
\centering
\includegraphics[width=1\linewidth]{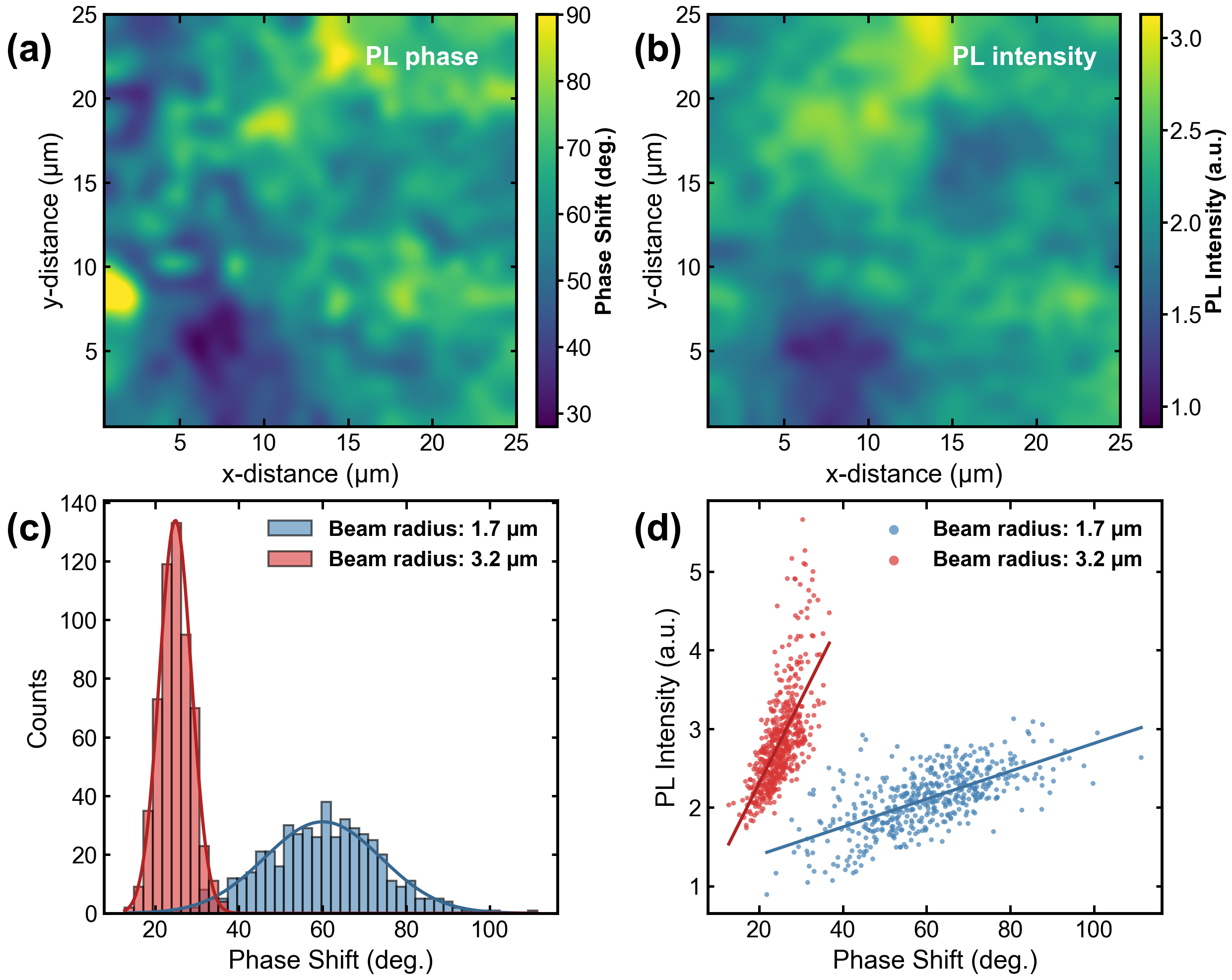}
\caption{(a) Phase shift map and (b) corresponding peak PL intensity (defined as the maximum AC and DC PL intensity signal, and taken as the PLQY proxy) over a 25 \textmu m $\times$ 25 \textmu m region (1 \textmu m step size) at $f =$ 50 mHz, collected using the custom-built pulsed laser microscopy setup. High-intensity (DC) excitation was provided by a focused 405 nm laser with a 1.7 \textmu m beam radius, while modulating excitation was provided using the 450 nm LED. Maps are bicubic-smoothed for visualization; corresponding raw data maps are shown in \textbf{Figure \ref{figSIRawMaps}}. (c) Phase histograms from the fluence-matched datasets, measured with laser beam radii of 1.7 \textmu m (blue) and of 3.2 \textmu m (red). (d) Scatter plots of the peak PL intensity versus phase shift for both beam sizes, with corresponding linear fits.}
\label{figMAP}
\end{figure}

To probe spatial variations across the perovskite film, we mapped the peak PL intensity -- the maximum AC and DC PL signal -- and the phase shift over a 25 \textmu m $\times$ 25 \textmu m area at $f =$ 50 mHz. These measurements were performed on a different setup than described earlier, in which pixels were acquired in random order to minimize interference between neighboring points. Specifically, this approach prevented photobrightening or phase segregation effects from altering signals in adjacent pixels\cite{photobright,phaseseg}. \\ Excitation was provided by a 405 nm pulsed laser (4 MHz repetition rate) at a fluence of 4.13 \textmu J/cm\textsuperscript{2}, focused to a beam radius of 1.7 \textmu m (see the SI for further details). The same 450 nm LED as before provided the sinusoidally modulating excitation. The resulting maps (\textbf{Figure \ref{figMAP}a} and \textbf{\ref{figMAP}b}) show strong spatial agreement between phase and PL intensity, further supporting that larger PL phase shifts coincide with higher PLQY.\\ 
To determine whether the beam size influences the IMPLS response, we mapped a different region of the same sample using a larger beam radius (3.2 \textmu m) while keeping fluence, modulation frequency and all other parameters constant. The full maps are shown in \textbf{Figure \ref{figSIRawMaps}}, while phase histograms comparing the two datasets are presented in \textbf{Figure \ref{figMAP}c}. Notably, the histograms differ in distribution, revealing a clear beam-size dependence of the phase response. This suggests that localized IMPLS measurements are sensitive to the excitation volume and potentially to transport processes -- such as ion diffusion -- rather than solely local properties. \\
Finally, scatter plots of PL intensity versus phase for both spot sizes (\textbf{Figure \ref{figMAP}d}) confirm the same positive correlation observed previously (\textbf{Figure \ref{fig4}d}), demonstrating that the general linear phase–PLQY relationship is robust across different excitation spot sizes. 

\section{Mechanistic Insights}\label{sec:mech}
\begin{figure}[!ht]
\centering
\includegraphics[width=0.7\linewidth]{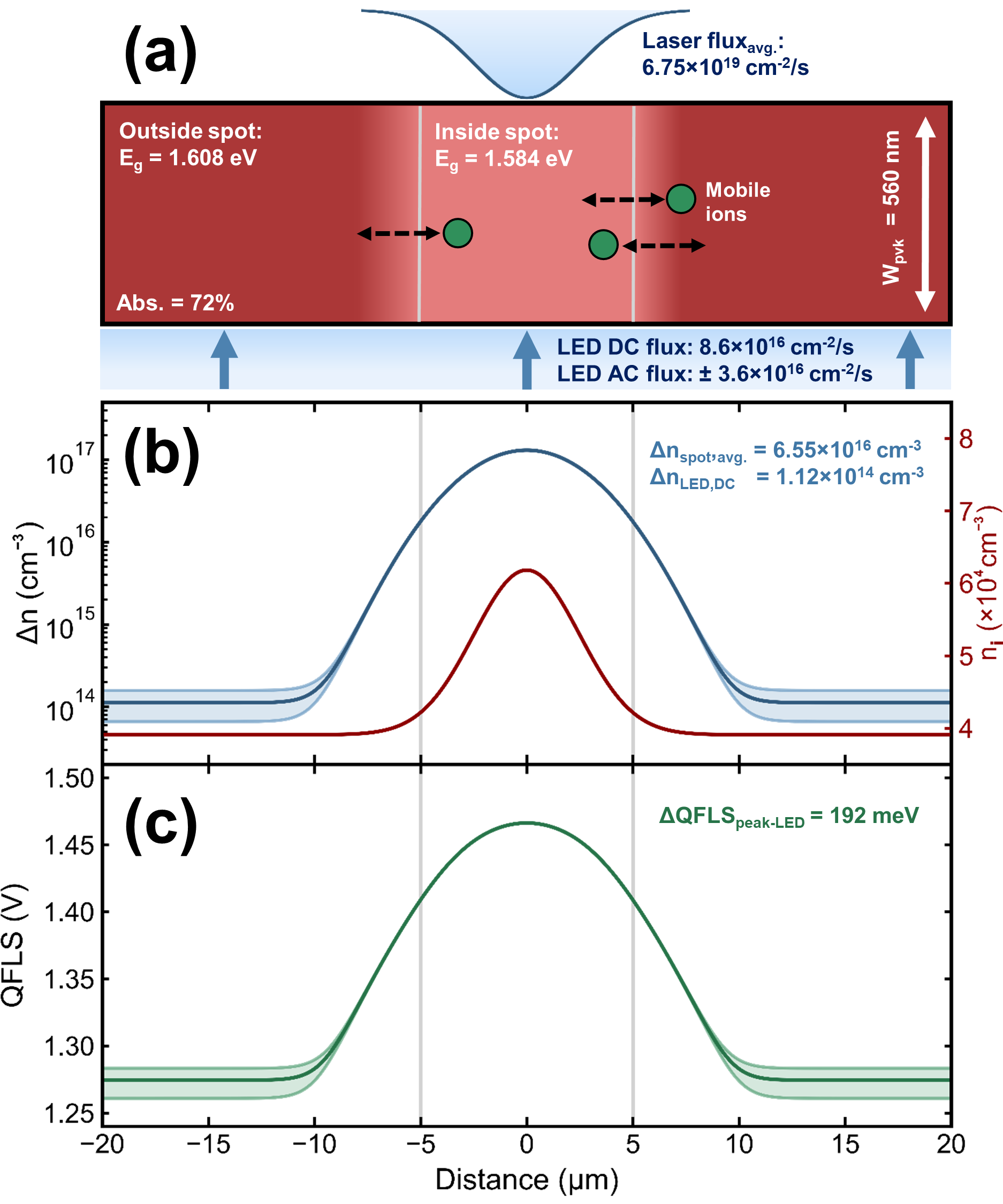}
\caption{(a) Schematic of perovskite film during a frequency-dependent IMPLS measurement. Different shades of red represent the different bandgap regions due to laser-induced halide segregation. Vertical gray lines indicate the 1/e\textsuperscript{2} diameter of the focused laser beam. Green markers with black arrows represent the sinusoidal mobile ion oscillations that likely occur along the beam edge and at sufficiently low modulation frequencies. (b) The minority excess carrier density (dark blue curve, left axis) and its amplitude (blue shaded region) across the same lateral distance as in panel (a). The intrinsic carrier density, $n_i$, is shown in red (right axis). (c) The DC and AC QFLS, calculated from the values in (b) and from the extracted doping density in the PLQY fitting analysis.}
\label{fig5}
\end{figure}

We now consider the underlying mechanism driving the IMPLS response observed in this work. In our previous IMPLS work, we attributed the slow process with \texttau\textsubscript{char,slow} $\approx$  77 s to iodide vacancy diffusion across the depth of the sample. Though we cannot perform a complete Nyquist analysis and generate (optical) equivalent circuits to extract a singular characteristic lifetime, we note that -- given the similar frequency range applied -- it is possible that we are observing similar diffusive processes of mobile ions in our response here. However, unlike in our previous work -- where the PL amplitude decreased with decreasing frequency -- we now observe a PL amplitude enhancement\cite{IMPLS}. To test whether this discrepancy arises from different excitation conditions, we repeated the frequency-dependent measurements without laser excitation and with the LED amplitude reduced to 20\% of its offset value. Under these conditions, the IMPLS trends are in agreement with our earlier findings. We previously attributed the decreasing PL amplitude to transverse (vertical) ionic diffusion toward the interface, leading to defect formation at the interface, thereby quenching the PL amplitude at low frequencies. In contrast, we propose that the PL enhancement observed under laser excitation (\textbf{Figure \ref{fig3}b} and \textbf{\ref{fig3}c}) is mediated by lateral ionic diffusion out of the beam spot\cite{photobright}. At modulation frequencies comparable to the characteristic diffusion time of the mobile ions, more ionic defects can escape laterally, reducing the local defect density and thereby enhancing the PL. This lateral diffusion process is illustrated schematically by the green markers in \textbf{Figure \ref{fig5}a}. This schematic is constructed using the beam size and excitation conditions applied in experiment (i) in this work.\\

A potential driver for ionic diffusion already proposed in the literature is due to asymmetric filling of electronic traps: if traps for one carrier remain filled longer than those of the other, a net electric field forms that decays radially with the beam-spot size\cite{controlling,photobright}. The field would then drive charged ionic defects out of the beam spot. An alternative, and arguably more straightforward, explanation is due to the order-of-magnitude higher hole mobility compared to electron mobility in perovskites, which can likewise generate an internal electric field that promotes ionic diffusion\cite{holemobility1,holemobility2}. \\ As a proxy for this electric field gradient, in \textbf{Figure \ref{fig5}b}, we plot $\Delta n$ (blue) across the same lateral distance as that in \textbf{Figure \ref{fig5}a}. Additionally, we plot intrinsic carrier density $n_i$ (red), where the change in $n_i$ is due to the bandgap difference between the iodide-rich region in the beam spot and the mixed-halide background. The quasi-Fermi level splitting (QFLS), calculated from $\Delta n$ and $n_i$, is shown in \textbf{Figure \ref{fig5}c}. Equations and calculation details for $n_i$ and the QFLS are listed in the SI. For these calculations, it is assumed that the charge carriers recombine close to where they were generated. If lateral diffusion of the electronic carriers were included, then the only difference would be a more gradual change in the $\Delta n$ and QFLS curves, but the general physics remains the same. An important aspect of consideration is that while the LED's modulation strongly affects $\Delta n$ and the QFLS outside the laser spot ($\Delta n$ and QFLS amplitudes are indicated by the blue and green shaded regions), the amplitude variations in $\Delta n$ and the QFLS within the 10 \textmu m beam diameter are negligible (changes are $<$0.1\%). This implies that mobile ions in the center of the beam spot are less likely to respond to the optical perturbation that would drive their oscillation; thus the IMPLS response is more likely to originate from the beam edge than from the center of the laser beam spot. \\

A notable point in this work is the PL phase shift and amplitude were still evolving even at the lowest frequency measured, corresponding to \texttau\textsubscript{meas. limit}  = 79.6 s. From literature, diffusion coefficients for ionic species, such as halide vacancies in mixed halide films, typically range between $D_{\textrm{ion}} = 10^{-9}-10^{-11}$ cm\textsuperscript{2}/s\cite{MAPIFutcher,McGovern2020,lucie2021,moritz1}. We can relate these coefficients to the lateral diffusion length, $L$, of the ionic species by\cite{Peng2018}:
\begin{equation}\label{eq:diffusion}
L = \sqrt{D_{\textrm{ion}} \textrm{\texttau}}
\end{equation}
In the simplest approximation, if we consider a mobile ion to migrate from the beam center to the region where the modulated excitation becomes relevant ($L\sim$10 \textmu m), the corresponding process time constant is estimated to range between 16 minutes and 27 hours ($f \approx$  1 \textmu Hz - 0.1 mHz). However, since the mobile ion distribution is not localized to the center of the spot, the averaged time constant will be lower. Though we cannot fit a complete optical equivalent circuit (OEC) model to the full IMPLS data to extract this time constant (since the DC term is also frequency-dependent), we can provide an estimate of the relaxation time by fitting only the phase data from \textbf{Figure \ref{fig3}a} to the non-ideal Cole-Cole expression:
\begin{equation}\label{eq:colecole}
\theta(f)
= \theta_0 +
\frac{A \, \sin\!\left(\frac{\alpha\pi}{2}\right)}
{\cosh\!\left[\alpha \ln\!\left(\frac{f}{f_\text{char}}\right)\right]
+ \cos\!\left(\frac{\alpha\pi}{2}\right)}
\end{equation}
where $f_\text{char}$ is the characteristic frequency, $\alpha$ is the non-ideality parameter, and $A$ and $\theta_0$ are amplitude and offset constants\cite{IMPLS,colecole}. With this expression, we yield $f_\text{char} = 3.2 \pm 0.95$ mHz (corresponding to \texttau\textsubscript{char} = $49.94 \pm 14.92$ s), and $\alpha = 0.6 \pm 0.1$ -- signifying a diffusive-like process\cite{tutorial}. The fit is shown in \textbf{Figure \ref{figSIcolecole}}. This low value would thus explain why we do not observe PL amplitude saturation or the expected turnover in phase as a function of frequency: the associated process relaxation frequency is at the limit of, or longer than, the measurement window. The hypothesis that mobile ions are diffusing laterally is further corroborated by the fact that the beam diameter of the laser directly influences the absolute phase shift, even under matched irradiance conditions (\textbf{Figure \ref{figMAP}c}). \\

We apply our hypothesis that lateral diffusion of mobile ions governs both the IMPLS and PLQY trends back to our original PLQY time series data in \textbf{Figure \ref{fig1}}. Accounting for the smaller beam diameter in the PLQY setup compared to the IMPLS setup and assuming the change in PLQY is due to ionic diffusive effects, then the corresponding diffusion coefficient for the time constant \texttau \textsubscript{1} = 42.1 s, is $D_\text{ion,1} = 1.52 \times10^{-10}$ cm\textsuperscript{2}/s. The slower diffusion coefficient for \texttau \textsubscript{2} = 664 s is $D_\text{ion,2} = 9.64 \times10^{-12}$ cm\textsuperscript{2}/s. Both of these diffusion coefficients are in agreement with the literature values for ionic species, providing further evidence to support our proposed mechanism driving the correlated IMPLS and PLQY data\cite{MAPIFutcher,lucie2021,moritz1}. This indicates that mobile ionic defects generally dominate the PLQY for this perovskite composition\cite{fenning}. In our earlier work, we additionally showed that the carrier lifetime is surface-limited, implying that interfacial mobile ionic defects are more influential than those in the bulk\cite{siliconinspired}. 

\section{Spatially Resolving Defect Types Across the Perovskite Film}
\begin{figure}[!ht]
\centering
\includegraphics[width=1\linewidth]{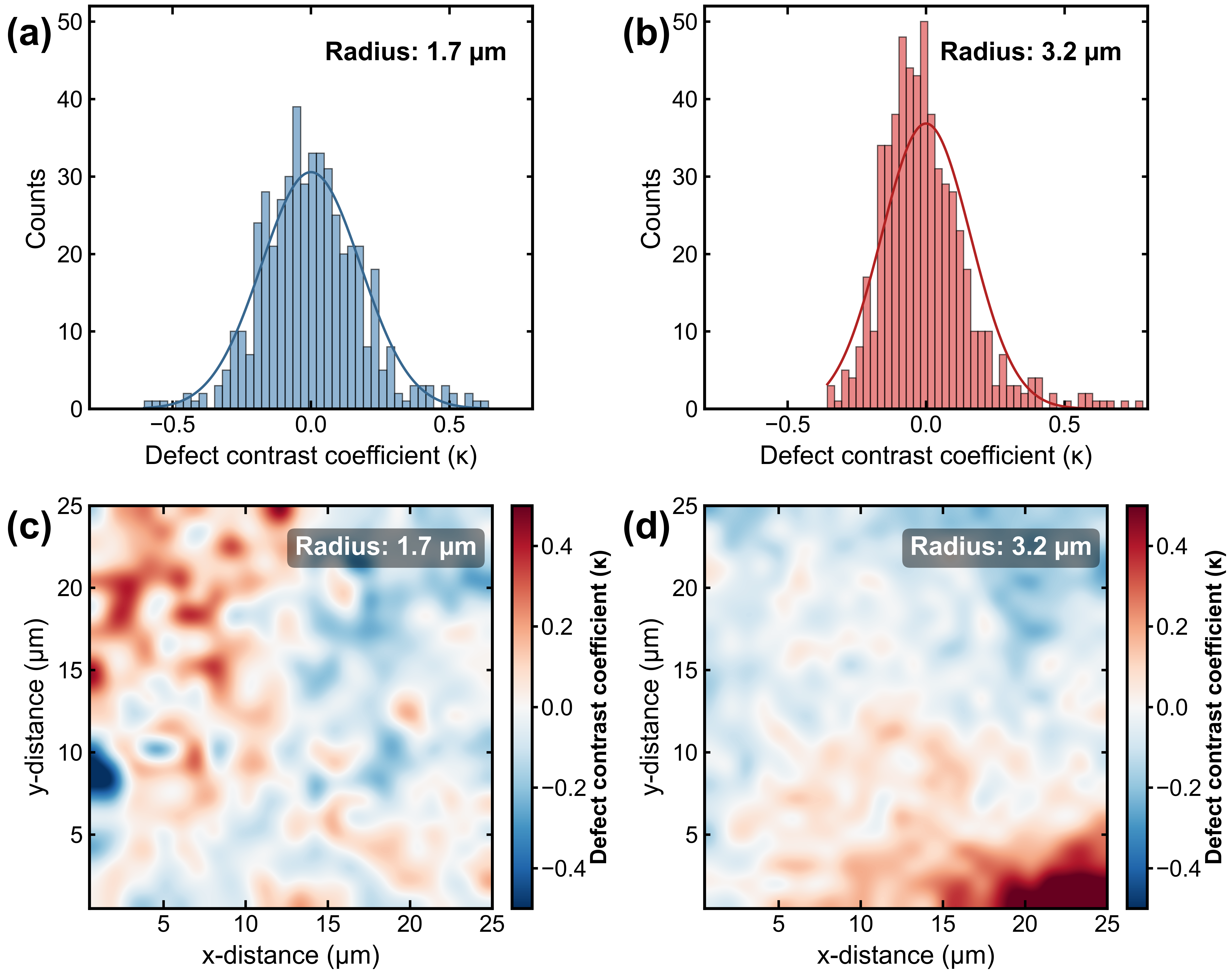}
\caption{Histograms of the defect contrast coefficient ($\kappa$) for the spatial data collected using the (a) 1.7 \textmu m and (b) 3.2 \textmu m beam size. Corresponding spatial maps of $\kappa$ for the same (c) 1.7 \textmu m and (d) 3.2 \textmu m radii, calculated using \textbf{Equation \ref{eq:DCC}} and shown after bicubic smoothing. Both maps exhibit notable spatial structure, while finer features are better resolved with the smaller 1.7 \textmu m beam size. From \textbf{Equation \ref{eq:DCC}}, regions with a positive DCC ($\kappa$ $>$ 0, red) indicate that the dominant defects are fixed (or highly immobile) at $f = 50$ mHz, whereas regions with a negative DCC ($\kappa$ $<$ 0, blue) correspond to areas where the dominant defect species are relatively more mobile.}
\label{fig7}
\end{figure}

Building on the logic that correlating IMPLS and PLQY analysis can reveal whether mobile ionic defects pin the PLQY, we apply this approach to identify deviations from this trend across the collected maps. In principle, this enables differentiation between regions where non-radiative recombination is limited by mobile ions versus regions dominated by fixed defects or by processes with characteristic frequencies far from the applied modulation frequency. \\To capture this behavior, we introduce a contrast parameter, $\kappa$, which we term the ``defect contrast coefficient'' (DCC). This parameter is defined as the difference between the normalized PL intensity (PLI) and the normalized phase:  
\begin{equation}\label{eq:DCC}
    \kappa (x,y) = \frac{\text{PLI}(x,y)}{{\text{PLI}}_\mu} - \frac{\theta (x,y)}{\theta_\mu}
\end{equation}
where $\text{PLI}_\mu$ and $\theta_\mu$ are the mean values over the mapped area. From this definition, regions with $\kappa > 0$ correspond to areas where the phase shift contribution is less significant than the PL intensity contribution, suggesting that mobile ionic responses at the applied frequency play a relatively smaller role in limiting the PL. In these regions, trap-assisted carrier recombination is more strongly governed by fixed or slower-timescale defects. Conversely, regions with $\kappa < 0$ indicate that mobile ionic defects in resonance with the applied modulation frequency dominate the recombination process. This distinction allows separation of the influence of mobile ions from that of fixed defects or much slower ionic species.\\ Using this definition, we determined $\kappa$ for both the 1.7 \textmu m and 3.2 \textmu m beam radii datasets. Histograms of $\kappa$ for both maps are shown in \textbf{Figure \ref{fig7}a} and \textbf{\ref{fig7}b}. The corresponding spatial maps of $\kappa$ for both beam radii are shown in \textbf{Figure \ref{fig7}c} and \textbf{\ref{fig7}d}, while the maps before smoothing are presented in \textbf{Figure \ref{figSIDCC}}. \\Notably, these maps exhibit spatial structure, suggesting that heterogeneity in defect types is resolvable using this approach. To statistically verify that the observed structure is not attributed to random noise, we calculated the global spatial autocorrelation coefficient using Moran's $I$\cite{Moran1950,MoranI}. Calculation details and further description of this approach are provided in the SI. For the map collected using the 1.7 \textmu m beam radius, $I = 0.325$, while $I = 0.566$ for the 3.2 \textmu m beam radius map. In both cases, $I > 0$ indicates statistically significant positive spatial autocorrelation, demonstrating that the $\kappa$ maps capture genuine spatial heterogeneity rather than spatial noise.\\

While this analysis is already useful for benchmarking sample quality and identifying dominant non-radiative pathways, it can also be extended to solar cell half-stacks to probe the influence of mobile ions at transport layer interfaces. We envision that frequency sweeps of this approach, combined with PL spectral mapping, could enable differentiation of mobile defect species: lower modulation frequencies would resolve slower ionic species (for example cation versus halide migration), while red-shifted PL regions would correlate with iodide-rich domains\cite{MAPIFutcher,McGovern2020,phaseseg}. Ultimately, this analysis provides a robust method to spatially resolve dominant loss pathways and distinguish competing recombination processes in perovskite films.

\clearpage

\section{Conclusion}
In conclusion, across the narrow frequency range of 2 mHz to 1 Hz and carrier densities of $\Delta n \approx 10^{14} - 10^{17}$ cm\textsuperscript{-3}, we observed that IMPLS phase and amplitude trends are directly linked to the sample's PLQY. We attribute these correlations to ionic defects diffusing laterally out of the high-intensity laser excitation spot. This interpretation is supported by the observed increase in PL amplitude at lower modulation frequencies, consistent with a reduction in the local trap-state density\cite{photobright}. Additional evidence comes from beam-size-dependent mapping experiments, which demonstrate that the IMPLS response depends on the area of the high-intensity excitation, and from control measurements in the absence of laser excitation which reproduce the PL amplitude quenching described in our earlier work\cite{IMPLS}. Together, these results indicate that mobile ionic defects -- likely located at the surface -- pin the PLQY across most regions of the perovskite film. \\
From a practical perspective, these findings suggest that lateral ion diffusion coefficients can be extracted either by extending IMPLS measurements across a broad frequency range or by systematically varying the laser beam size. Moreover, under the specific conditions where PLQY is limited by mobile ions, the phase response itself can serve as a direct proxy for material quality\cite{fenning}. Spatial discrepancies between the phase response and PLQY thus provide a means to distinguish regions where non-radiative recombination is dominated by mobile ions (corresponding to a negative DCC, blue regions across the spatial maps in \textbf{Figure \ref{fig7}}) from areas dominated by fixed or slower defects (corresponding to a positive DCC, red regions in \textbf{Figure \ref{fig7}}). \\
As an outlook, combining frequency- and beam-size-dependent IMPLS with spectral analysis could yield deeper insight into the specific defect species governing the response. With further benchmarking and complementary methods -- such as direct ion-density quantification with drift–diffusion simulations -- we aim to establish a comprehensive model of the coupled electronic–ionic processes underlying IMPLS in the near future\cite{lucie2021,moritz1,moritz2,futscher2020,hauffis,Aguscomb}.



\medskip

\medskip
\textbf{Acknowledgements} \par 
The work is part of the Dutch Research Council (NWO) in collaboration between AMOLF and TNO. The work was performed at the NWO institute AMOLF. E.C.G., J.G., A.O.A., and B.E. received funding from the European Research Council (ERC) under the European Union's Horizon Europe research and innovation programme (grant agreement no. 101043783 for E.C.G. and J.G., grant agreement no. 947221 for A.O.A. and B.E.). This work is partly funded through governmental funding of TNO financed by the Ministry of Climate Policy and Green Growth and Ministry of Economic Affairs.\clearpage

\setcounter{table}{0}
\renewcommand{\thetable}{S\arabic{table}}%
\setcounter{figure}{0}
\renewcommand{\thefigure}{S\arabic{figure}}%
\setcounter{equation}{0}
\renewcommand{\theequation}{S\arabic{equation}}
 


\title{\textbf{Supporting Information}}

\maketitle


\author{}

\begin{affiliations}
\end{affiliations}




\section{Sample Fabrication}
In a nitrogen-filled glovebox, two 1.5 M solutions of \ce{PbI2} ($\geq$99.99\%, TCI) and \ce{PbBr2} ($\geq$98\%, TCI) were prepared in DMF:DMSO (4:1 by volume) and stirred overnight at 70 $^\circ$C. The \ce{PbI2} solution was combined with \ce{FAI} powder ($\geq$99\%, TCI) and excess DMF:DMSO solvent to yield a 1.24 M \ce{FAPbI3} solution containing 10 mol\% excess \ce{PbI2}. Similarly, the \ce{PbBr2} solution was combined with MABr ($\geq$99\%, TCI) to form a 1.24 M \ce{MAPbBr3} solution, also with a 10 mol\% \ce{PbBr2} excess. Both perovskite solutions were stirred for an additional 2 hours at 70 $^\circ$C. The solutions were then mixed in an 80:20 ratio (\ce{FAPbI3}:\ce{MAPbBr3}), followed by the addition of a 1.5 M \ce{CsI} solution in DMSO ($\geq$99.99\%, Sigma-Aldrich) to achieve the final composition \ce{Cs_{0.07}(FA_{0.8}MA_{0.2})_{0.93}Pb(I_{0.8}Br_{0.2})3}. The resulting solution was stirred at 70 $^\circ$C for another 2 hours before cooling to room temperature and filtering through a 0.45 \textmu m PTFE filter. \\
Glass substrates were cleaned by scrubbing them with a 1\% Hellmanex III solution in deionized (DI) water, followed by applying sequential sonication steps (15 minutes each) in 70 $^\circ$C water, acetone, and finally in isopropanol. Immediately before perovskite spin-coating, the substrates were treated under UV-ozone for 30 minutes and then transferred into the glovebox. \\
120 \textmu L of the precursor solution was dispensed onto the glass substrate and subsequently spin-coated at 5000 RPM for 30 seconds after a 6 second ramp-up time. At 15 seconds before the end of the cycle, 170 \textmu L of chlorobenzene filtered through a 0.22 \textmu m PTFE filter was deposited as an antisolvent. The films were annealed on a hotplate at 100 $^\circ$C for 45 minutes. Finally, the films were encapsulated with 60 nm of \ce{SiO2}, deposited by electron-beam evaporation (Polyteknik Flextura M508E), directly from a \ce{SiO2} target at a deposition rate of 0.06 nm/s.

\section{Measurement Details}\label{sec:SIDiffLT}
The IMPLS setup shown in \textbf{Figure \ref{fig2}a} of the main text is comprised of a WITec alpha300 SR confocal imaging microscope, coupled to a S1DC405 405 nm CW diode laser (ThorLabs) and to a 450 nm light-emitting diode (Cree LED). The LED intensity and sinusoidal modulation was driven with a 33522A arbitrary waveform generator (Keysight Technologies). The 405 nm focused beam size was measured using the knife-edge technique and the power density and equivalent photon flux were determined with a wavelength-calibrated power meter (S120VC coupled to a PM100D, ThorLabs)\cite{blade}. \\
IMPLS maps (\textbf{Figures \ref{figMAP}}, \textbf{Figure \ref{figSIRawMaps}}) and uncertainty statistics (\textbf{Figure \ref{figSIhistogram}}) were obtained using a home-built pulsed laser microscopy setup. A tunable laser (Chameleon Discovery NX, Coherent) set to 405 nm was coupled to a pulse picker (pulseSelect, APE) to produce pulses (80 fs pulse width) at a repetition rate of 4 MHz. The beam was spatially filtered and then focused onto the sample using a Mitutoyo M Plan APO NIR B $\times$50 objective. The size of the beam on the sample was adjusted by changing the focal length of the collimating lens after the spatial filter, and the spot size was calibrated using the knife-edge technique. The sample was mounted on a TRITOR 100 SG piezo stage (Piezosystem Jena), and the same 450 nm LED as above provided sinusoidal AC illumination from the rear of the perovskite sample. To collect the maps, the measurement sequence was randomized, with each consecutive point separated by at least 5 \textmu m from the previous measurement. The total PL signal was collected using a Kymera 193i (Andor) and a ProEM electron-multiplying CCD (Princeton Instruments). \linebreak

\begin{figure}[H]
\centering
\includegraphics[width=1\linewidth]{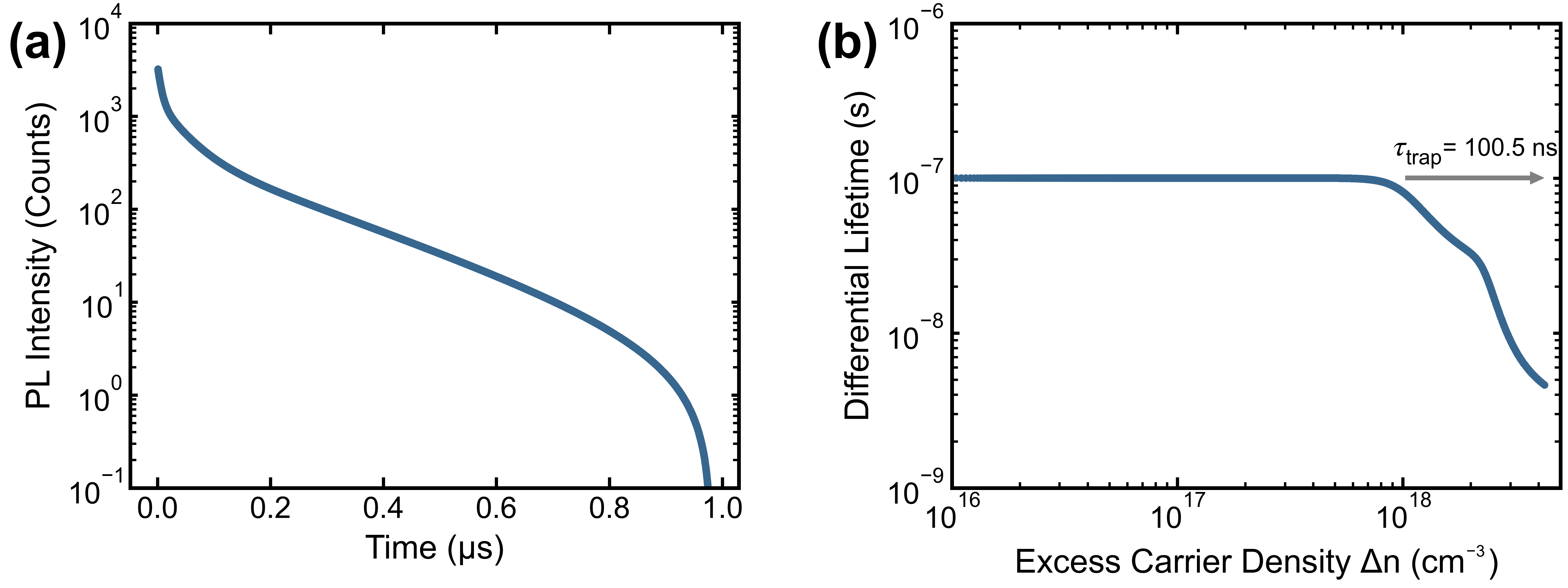}
\caption{(a) Arbitrary fit of the averaged raw PL decay trace of the encapsulated perovskite thin film sample, following noise floor signal subtraction. (b) Differential lifetime as a function of excess minority carrier density, $\Delta n$. The saturation point is indicated with the gray arrow, in which $\textrm{\texttau}_{\textrm{diff}} = \textrm{\texttau}_{\textrm{trap}}$ = 100.5 ns\cite{underkirchartz}.}
\label{figSI1}
\end{figure}

The time-resolved photoluminescence decay trace for the thin film was obtained with a time-correlated single-photon counting (TCSPC) system (PicoQuant), comprised of a PDL 828 Sepia II, a HydraHarp, a 485 nm pulsed diode laser and an Olympus $\times$60 Plan Apochromat water objective. The TRPL decay was measured following approximately 2 minutes of continuous light soaking. The total decay was averaged across an 80 \textmu m$\times$80 \textmu m map with a 400 nm step size between each pixel. The laser repetition rate was set to 1 MHz, sufficiently low to capture the entire PL decay\cite{kirchartzshallow}. An arbitrary fit was applied to the TRPL signal to reduce noise, before the fit was background corrected by subtracting the noise floor from the signal. This decay is shown in \textbf{Figure \ref{figSI1}a}. From measuring the film thickness by profilometry, absorptance, laser fluence (0.136 mJ/cm\textsuperscript{2}) and the laser spot size (1/e\textsuperscript{2} diameter = 2 \textmu m), the excess minority carrier density generated from each pulse was determined, $\Delta n_0$ = 4.248$\times$10\textsuperscript{18} cm\textsuperscript{-3}. From this starting carrier density, the complete PL decay trace was converted to the decay of excess carrier density using $\Delta n \propto \sqrt{\mathrm{PL}}$. Then, the differential lifetime was calculated as a function of $\Delta n$ with the differential lifetime equation:
\begin{equation}\label{eq:diff}
    \textrm{\texttau}_{\textrm{diff}}(\Delta n) = - \Big( \frac{\textrm{d}\ln \Delta n(t)}{\textrm{d}t}\Big)^{-1}
\end{equation}
By definition, the saturation point of this curve is where $\textrm{\texttau}_{\textrm{diff}} = \textrm{\texttau}_{\textrm{trap}}$ = 100.5 ns (\textbf{Figure \ref{figSI1}b})\cite{underkirchartz}. \\

Photoluminescence quantum yield maps were collected using a custom-built integrating sphere microscopy setup and software (separate to the system described for IMPLS mapping above); a schematic is shown in our previous work\cite{mann2016}. 660 nm excitation was provided using a 78 MHz supercontinuum laser (NKT Fianium FIU 15) coupled to an acousto-optical tunable filter (AOTF, with a Gooch \& Housego AODS20200-8 driver). The AOTF was programmed to select the appropriate excitation wavelength, intensity and to modulate the laser light based on the input reference frequency. The beam radius was measured to be 800 nm using the knife-edge technique. Two PDA100A Si photodetectors were used as a beam monitor (BM) detector and as a reflection (R) detector. The BM detector collected the light emitted directly from the laser. The R detector collected light that was directly reflected or emitted within the acceptance angle of the objective lens from the integrating sphere after sample interaction. A calibrated Newport 818-UV photodetector was used to collect the transmitted and scattered light from the integrating sphere (IS), and to convert the signal intensity to photon counts. All detectors were connected to three lock-in amplifiers (LIA, Stanford Research Systems SR830). One LIA was used to set the reference frequency, which was transmitted to the AOTF and to the other LIAs. The reference frequency was set to 1.253 kHz. Four reference measurements were performed: a blank to calibrate the PDA100A detectors, a filter leakage reference, a 100\% transmission reference, and a 100\% reflection reference. The reflection reference was measured with a silver coated mirror placed at the same position as the sample during the PLQY measurement (ThorLabs PF10-03-P01). 700 nm short-pass and long-pass filters (placed in front of the reflection and IS detectors) were used to separate the absorptance and PL data, respectively. In addition to selecting the laser intensity using the AOTF, ND filters were applied to sweep the intensity range. The sample was mounted on an external piezo stage while in the integrating sphere, enabling the 5 \textmu m$\times$5 \textmu m PLQY maps to be collected. \clearpage
\begin{figure}[!ht]
\centering
\includegraphics[width=1\linewidth]{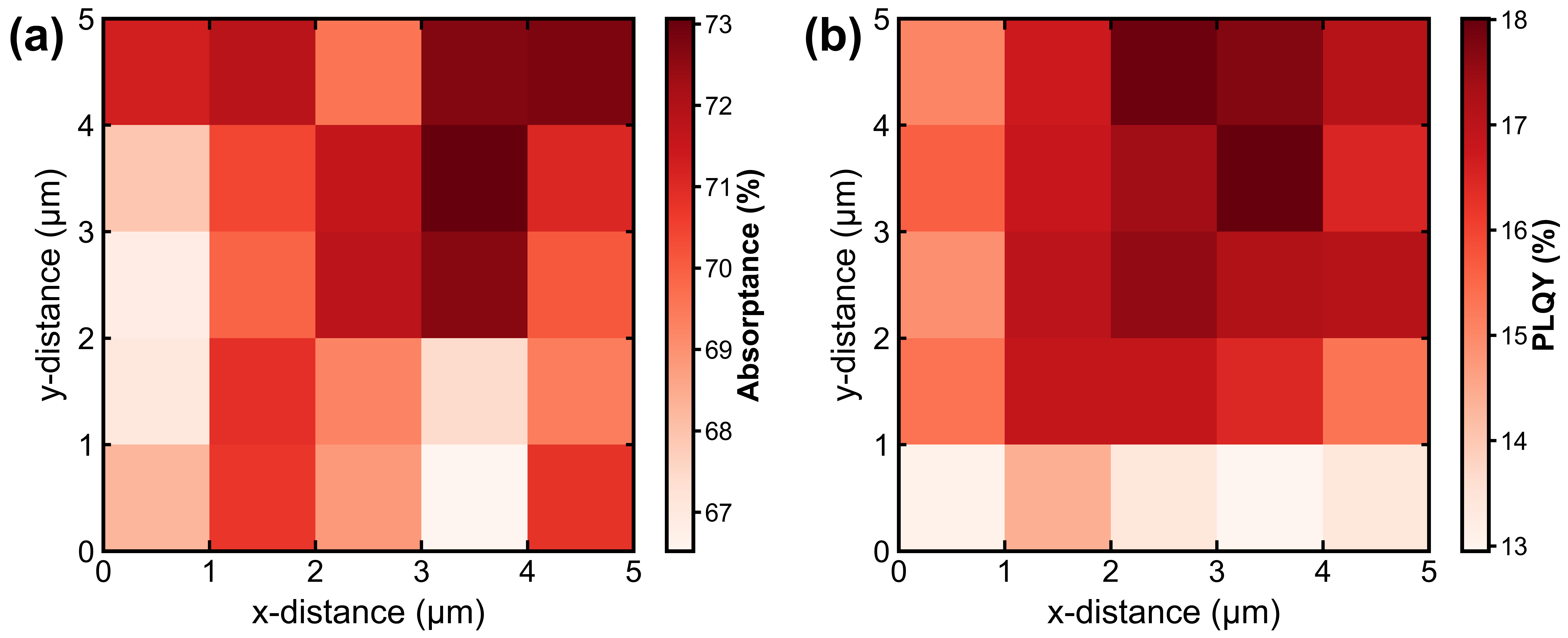}
\caption{(a) Absorptance map of the perovskite thin film measured under an incident photon flux of $\Phi_\text{exc} = 5.88\times10^{20}$ cm\textsuperscript{-2}/s. Within this map, the average absorptance and standard deviation are $A=70.1\%$ and $\sigma_A = 1.9\%$. (b) Corresponding PLQY map for the same area. The average PLQY and standard deviation within the map are PLQY$=16.0\%$ and $\sigma_{\textrm{PLQY}} = 1.6\%$.}
\label{figSI2}
\end{figure}

For every measurement, the emission data was obtained first, then the absorptance ($A$) data was collected. After calibration to convert the raw detector signals to number of photons ($\Phi$) and accounting for the reference measurements, the PLQY was determined using the standard:
\begin{equation}\label{eq:PLQYSI}
    \textrm{PLQY} = \frac{\Phi_{PL}}{\Phi_\text{abs}} = \frac{\Phi_{PL,R} + \Phi_{PL,IS}}{\Phi_\text{exc}\times A}
\end{equation}
where $\Phi_{PL,R}$ and $\Phi_{PL,IS}$ represent the number of emitted photons incident on the R detector and the IS detector. The IS detector was used to collect both transmission, T, and scattering, S. With the short-pass filters before the detectors, the absorptance was calculated using $A = 1 - R - S - T$.
Exemplary maps of the absorptance and PLQY are shown in \textbf{Figure \ref{figSI2}}. \\

\section{Supporting IMPLS Characterization}\label{sec:SIIMPLSchar}
\begin{figure}[H]
\centering
\includegraphics[width=0.65\linewidth]{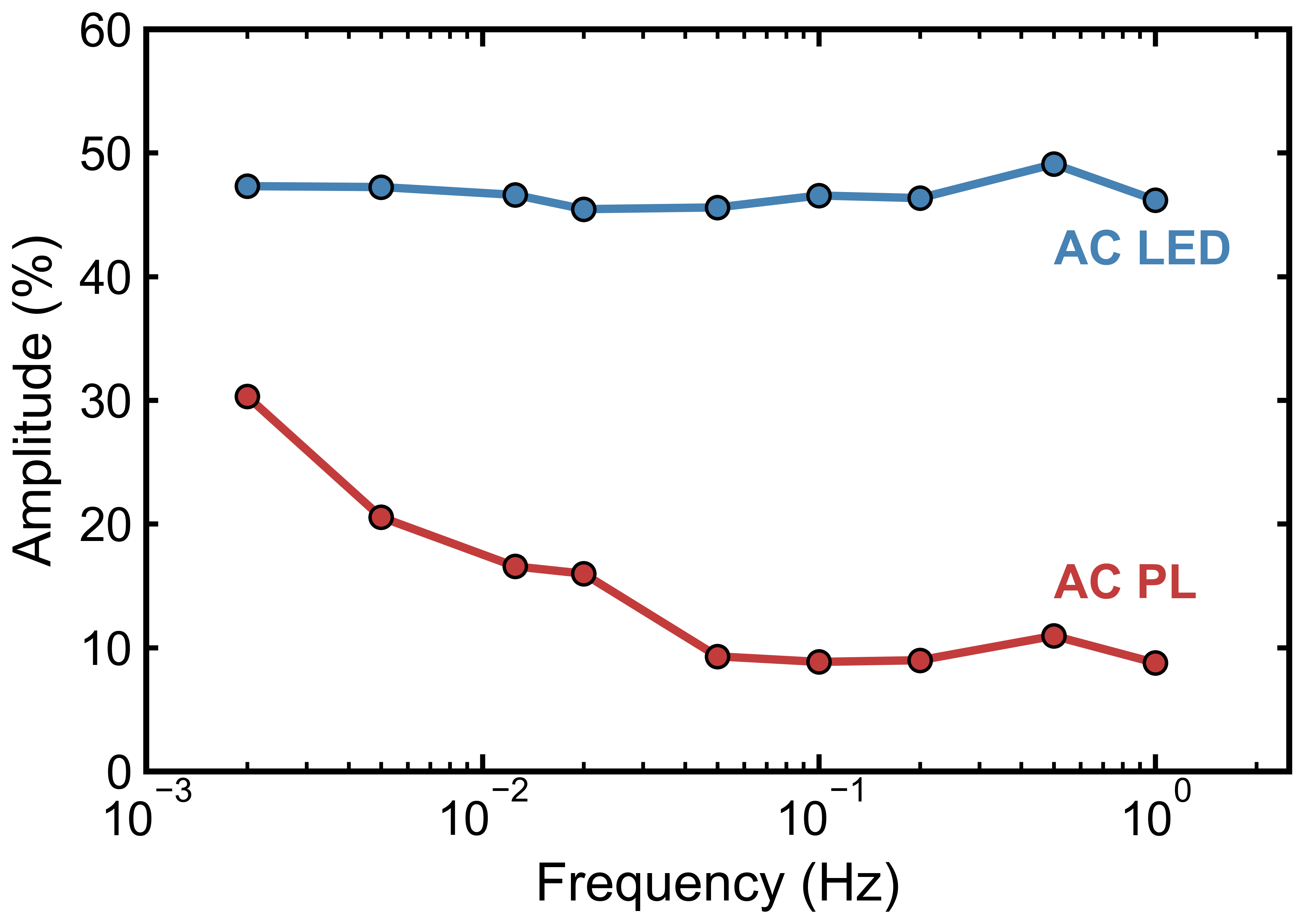}
\caption{Comparison between the LED amplitude (blue) and the PL amplitude (red) as a function of frequency. The LED amplitude is independent of frequency, signifying that the change in the PL amplitude is due to the sample-related processes.}
\label{figSI3}
\end{figure}

To maximize the signal-to-noise ratio across IMPLS measurements with different frequencies, different integration times were applied to acquire the spectra. To account for the differences in the integration times, the PL DC offset was scaled relative to the LED offset in each spectrum, as the absolute LED offset was constant for all measurements. Thus:
\begin{equation}
{PL}(f)_{\textrm{DC,relative}} = \frac{PL(f)_{\textrm{DC,counts}}}{LED(f)_\textrm{{DC,counts}}}
\end{equation}
Where counts represent the integrated CCD counts for each spectrum. The PL amplitude was defined as the ratio between the PL peak value and the PL offset. Typically for other modulated techniques at high frequencies, the response amplitude is also scaled by the amplitude of the input parameter to correct for any frequency-dependent system response. However, given the low modulation frequency range explored in this work, the LED amplitude did not vary as a function of frequency (\textbf{Figure \ref{figSI3}}) and so the PL amplitude did not need to be corrected further.

As noted in the main text, all frequency measurements for the frequency sweep in \textbf{Figure \ref{fig3}} were performed in random order and always on a new spot on the sample to prevent any long-term or frequency-induced changes from altering the IMPLS response. Further, the PL data was only fitted to a sinusoidal function after the DC PL signal stabilized. In other words, for every frequency point, the DC PL data remained constant across the measurement. This ensured that the observed changes in the PL offset signal in \textbf{Figure \ref{fig3}c} are due to probed ionic process, and not from any external factor such as heating, degradation, or other frequency-independent chemical reactions.\linebreak

\begin{figure}[H]
\centering
\includegraphics[width=0.9\linewidth]{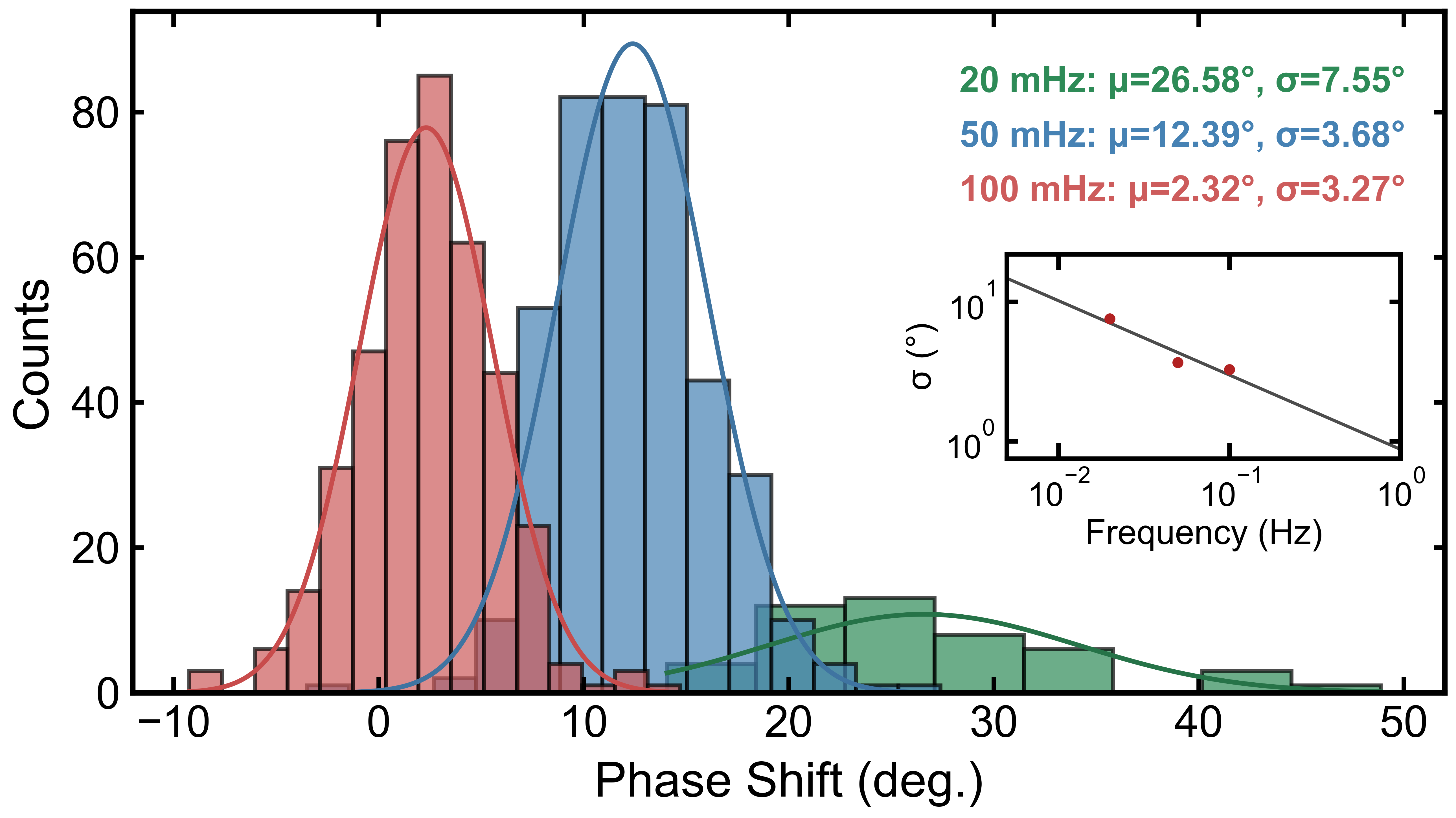}
\caption{Histograms of the phase shift distributions measured at three modulation frequencies: $f$ = 100 mHz (red), 50 mHz (blue) and 20 mHz (green). Solid curves indicate Gaussian fits to the data, from which the mean ($\mu$) and standard deviation ($\sigma$) were extracted. The inset shows the frequency dependence of $\sigma$, with the dark gray line indicating the fit.}
\label{figSIhistogram}
\end{figure}

\begin{figure}[H]
\centering
\includegraphics[width=1\linewidth]{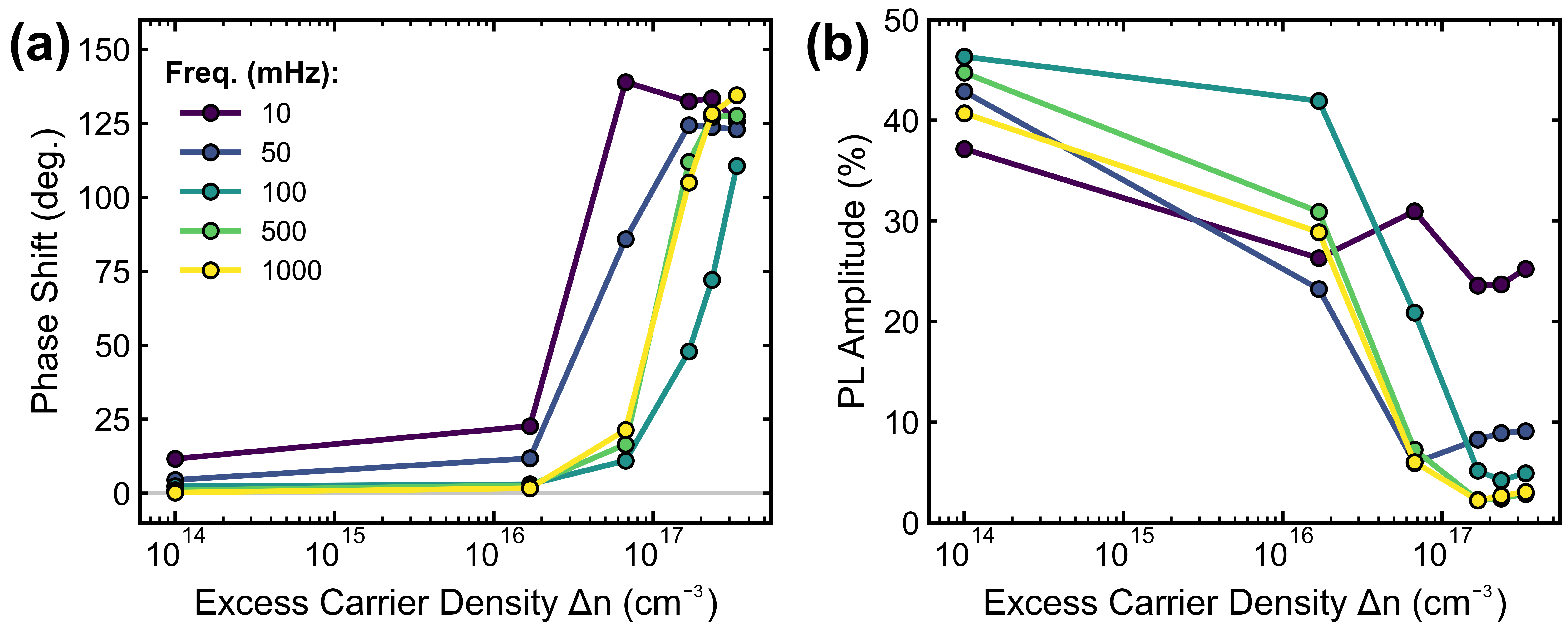}
\caption{Line scans of the (a) PL phase shift and (b) PL amplitude as functions of $\Delta n$. Each color represents a different frequency applied during the measurement, labeled in the top left corner in panel (a).}
\label{figSI5}
\end{figure}

To estimate the uncertainty arising from both the fitting procedure and sample inhomogeneity, we performed repeated measurements at three modulation frequencies ($f =$ 100, 50, 20 mHz) across different points on the sample. For the 100 mHz and 50 mHz cases, $N= 400$ data points were collected, while for the 20 mHz case only $N= 47$ data points were obtained due to the longer measurement time required. The resulting histograms of the phase shift distributions are shown in \textbf{Figure \ref{figSIhistogram}}. At all three frequencies, the distributions are well described by Gaussian statistics (solid curves). As illustrated by the inset, the extracted standard deviation ($\sigma$) of the fits decreases with increasing frequency. The relative error, defined as RE = 100\%$ \times \sigma/( \mu\sqrt{N})$ with respect to the mean ($\mu$), ranged between 1.5\% and 7\% for the three measured frequencies. The offset in the absolute phase values relative to \textbf{Figure \ref{fig3}a} arises from the different excitation conditions and beam size; these statistical measurements were collected using the custom-built ultrafast laser setup with 405 nm pulsed laser excitation.

\begin{figure}[H]
\centering
\includegraphics[width=0.9\linewidth]{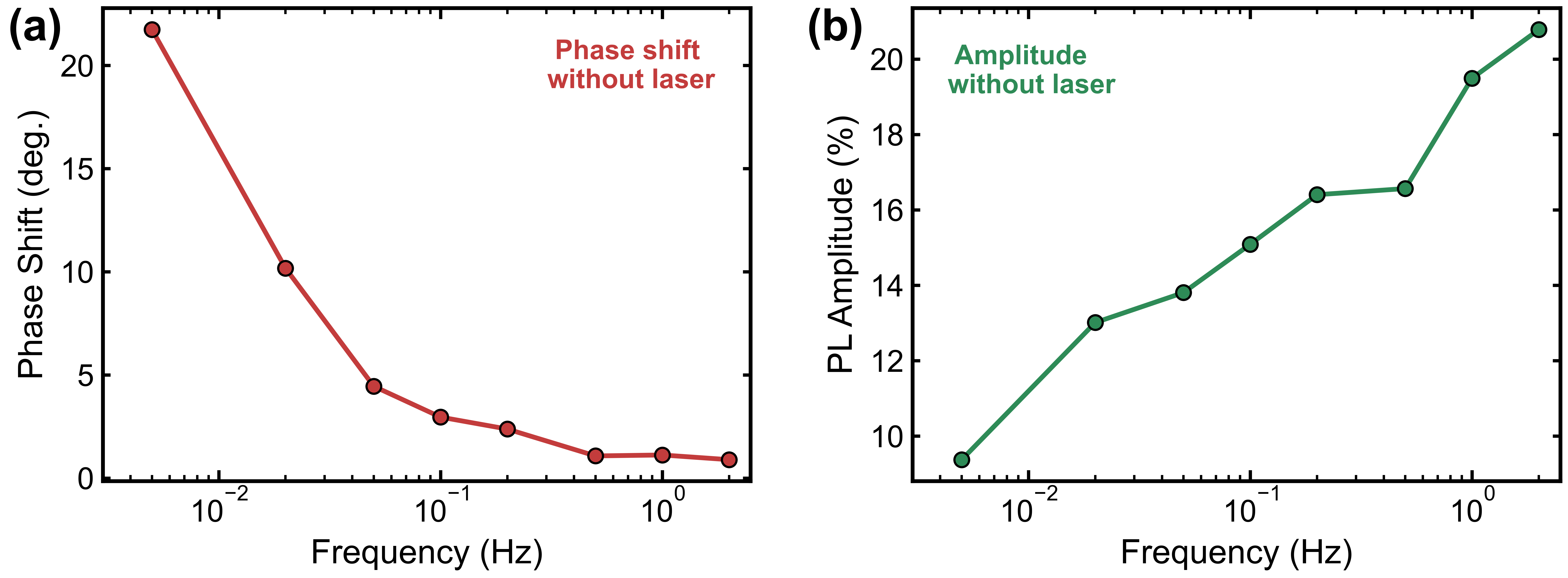}
\caption{(a) PL phase shift and (b) PL amplitude as functions of modulating frequency without the additional laser excitation. For this measurement, the LED amplitude was reduced to 20\% of its offset value. The sample's amplitude is reduced as frequency is reduced and the extent of the phase shift is significantly reduced without the additional laser excitation, compared to \textbf{Figure \ref{fig3}}. These trends in panels (a) and (b) agree with those observed for a similar perovskite composition in our previous work\cite{IMPLS}.}
\label{figSI4}
\end{figure} 

\begin{table}[htbp]
\centering
\caption{Parameters obtained from fitting $\theta$($\Delta n$) to the logarithmically-scaled Gaussian model (\textbf{Equation \ref{eq:logspaceG}}). The associated uncertainties of this empirical model were determined directly from nonlinear least-squares analysis.}
\begin{tabular}{ccccc}
\toprule
\(\bm{A}\) \textbf{(deg.)} & 
\(\bm{B}\) \textbf{(deg.)} & 
\(\bm{\mu}\) \textbf{(\(\text{cm}^{-3}\))} & 
\(\bm{\mu_{\log}}\) & 
\(\bm{\sigma_{\log}}\) \\
\midrule
\(111.1 \pm 4.5\) & \(15.81 \pm 3.3\) & \((1.88 \pm 0.16)\times10^{17}\) & \(17.27 \pm 0.04\) & \(0.398 \pm 0.038\) \\
\bottomrule
\end{tabular}
\label{tab:gaussian_params}
\end{table}

\begin{figure}[H]
\centering
\includegraphics[width=0.65\linewidth]{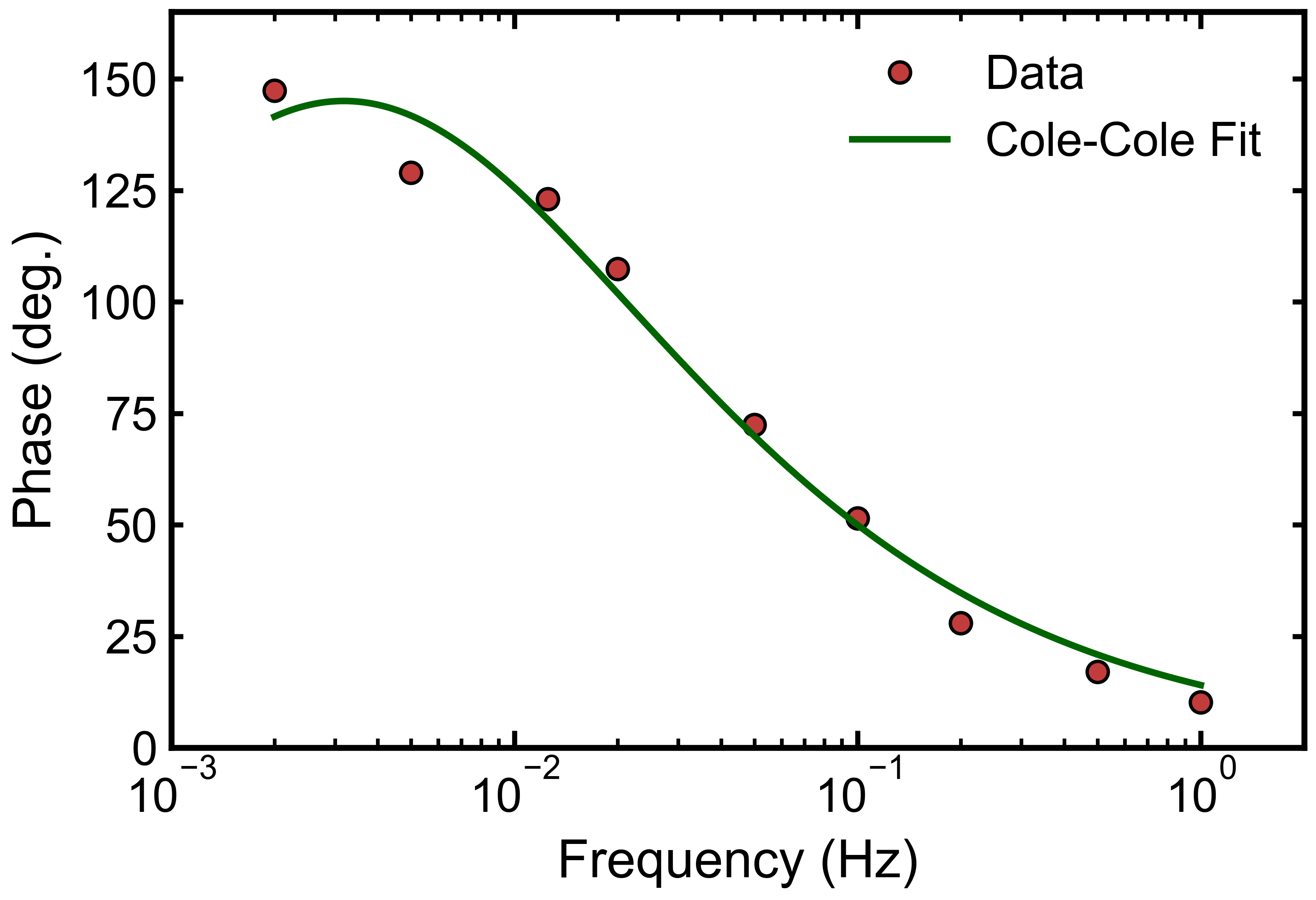}
\caption{Phase shift data from \textbf{Figure \ref{fig3}} of the main text (red), and the Cole-Cole fit to the data using \textbf{Equation \ref{eq:colecole}} (green).}
\label{figSIcolecole}
\end{figure} 

In \textbf{Figure \ref{fig5}b}, the intrinsic carrier density $n_i$ was calculated by approximating the perovskite band gap as the peak energy position of the PL with the laser removed (background band gap) and with the laser applied (laser region). From the band gap, the dark recombination current density at the radiative limit, $J_{0,rad}$ was calculated from detailed balance principles, and $n_i$ was solved using\cite{dblim,ordersofrec}:
\begin{equation}
    n_i = \sqrt{\frac{J_{0,rad}}{q\times W_{pvk}\times k_{rad}}}
\end{equation}
In \textbf{Figure \ref{fig5}c}, the QFLS was then determined using the standard expression:
\begin{equation}
    QFLS = \frac{kT}{q}\ln\Big( \frac{np}{n_i^2}\Big) \approx \frac{kT}{q}\ln\Big( \frac{n[p_0+n]}{n_i^2}\Big)
\end{equation}

\section{Supporting IMPLS Maps Analysis}\label{sec:SIIMPLSmaps}
\begin{figure}[H]
\centering
\includegraphics[width=1\linewidth]{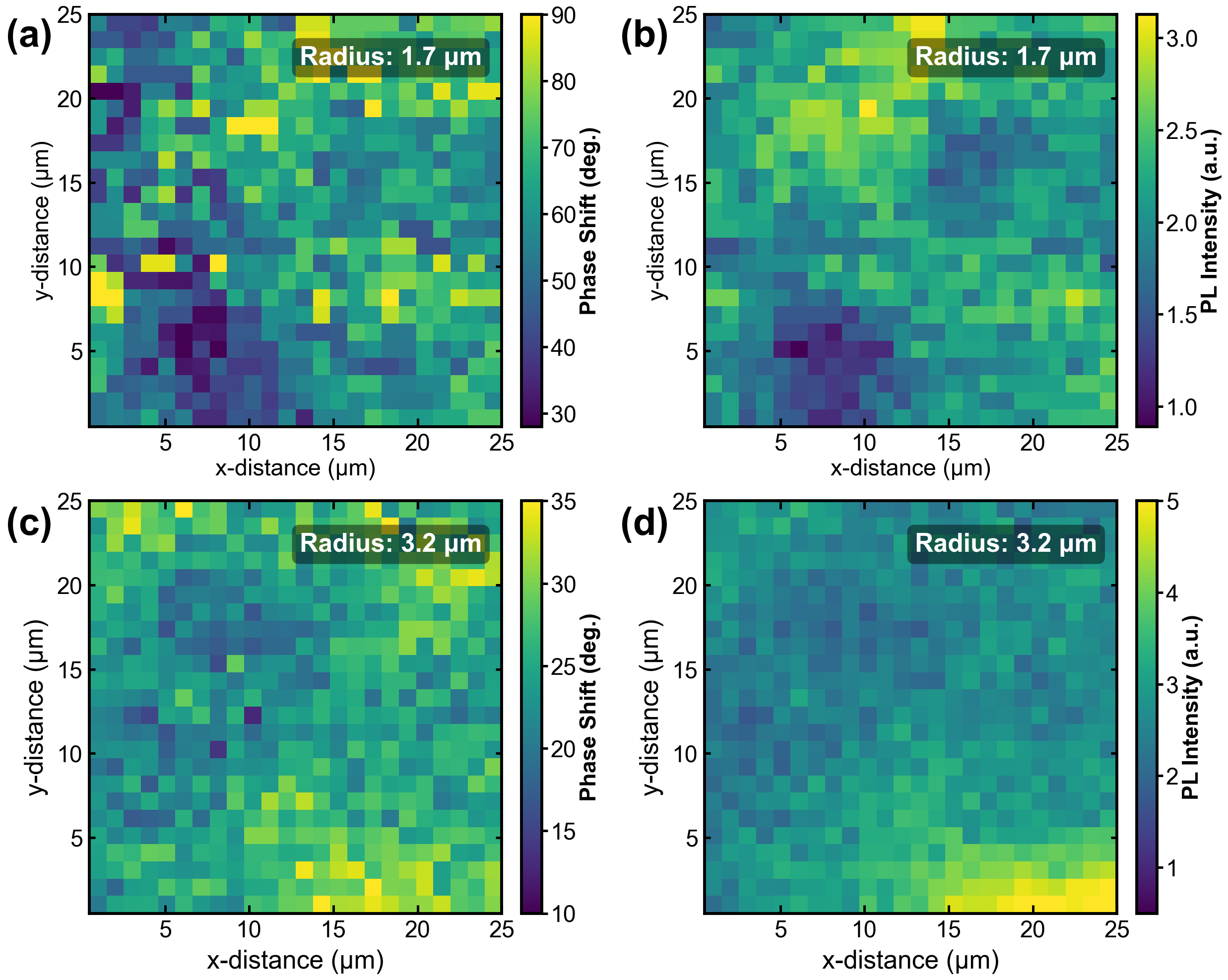}
\caption{(a) Phase shift and (b) PL intensity raw signal maps (before bicubic smoothing) for a beam radius of 1.7 \textmu m. (c) Phase shift and (d) PL intensity raw signal maps for a beam radius of 3.2 \textmu m. The 3.2 \textmu m measurement was performed on a different area of the same sample, at matched fluence and with the same modulation frequency ($f$ = 50 mHz).}
\label{figSIRawMaps}
\end{figure}\clearpage

\begin{figure}[!ht]
\centering
\includegraphics[width=1\linewidth]{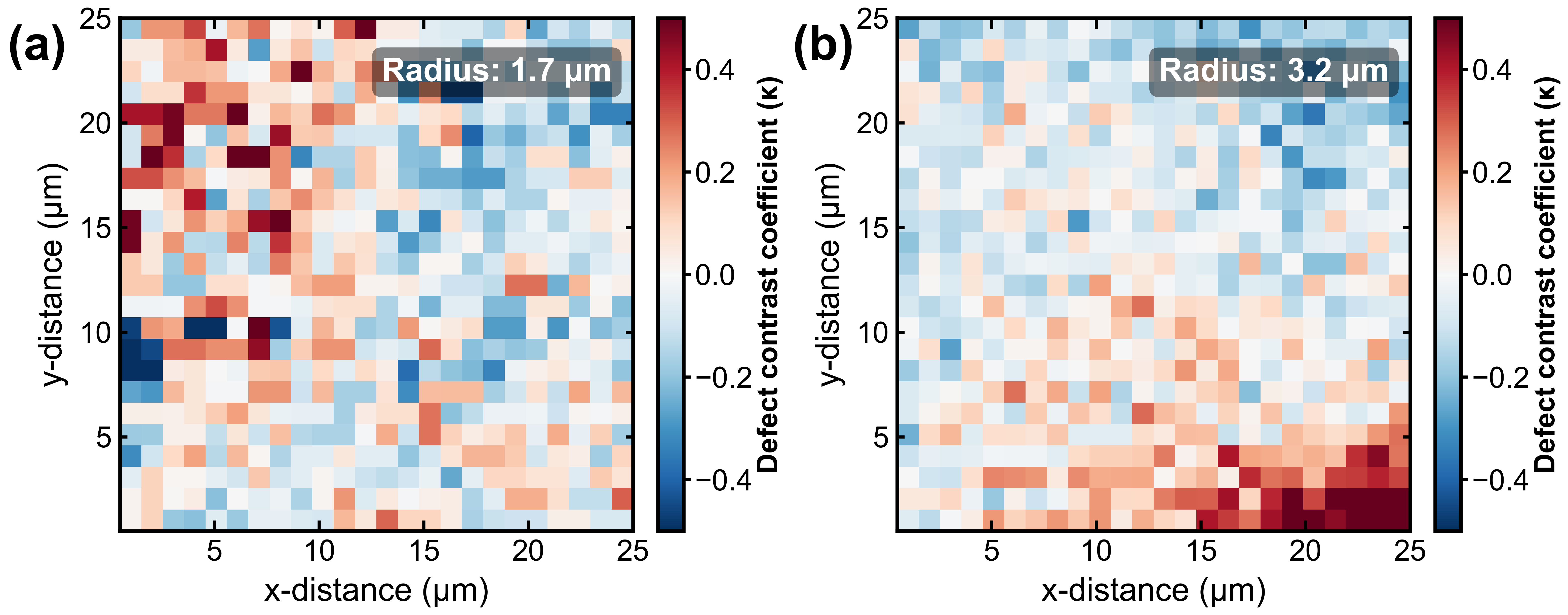}
\caption{Defect contrast coefficient ($\kappa$) maps obtained using \textbf{Equation \ref{eq:DCC}} for beam radii of (a) 1.7 \textmu m and (b) 3.2 \textmu m, prior to bicubic smoothing.}
\label{figSIDCC}
\end{figure}

To quantify whether the observed $\kappa$ maps exhibit spatial structure beyond random noise, we calculated the global Moran's $I$ statistic for spatial autocorrelation\cite{Moran1950}. Moran's $I$ quantifies spatial autocorrelation on a scale from $I = -1$ (perfect dispersion) to $I =1$ (perfect clustering), with $I = 0$ corresponding to spatial randomness. While Moran's $I$ is widely used in geographical applications, it has recently also been applied to analyze chemical disorder with scanning transmission electron microscopy (STEM) maps\cite{MoranI}. Moran's $I$ for a map containing $N$ pixels is defined as:
\begin{equation}
I = \frac{N}{\sum_{i}\sum_{j} w_{ij}} \,
    \frac{\sum_{i}\sum_{j} w_{ij} (x_i - \bar{x})(x_j - \bar{x})}
         {\sum_{i} (x_i - \bar{x})^2},
\end{equation}
where $x_i$ is the value at pixel $i$, $\bar{x}$ is the mean value over all pixels, and $w_{ij}$ is the spatial weight between pixels $i$ and $j$. Specifically, $w_{ij}$ determines whether a pair of pixels is considered close enough to contribute to the measure of spatial autocorrelation. Here, we adopted a nearest-neighbor weighting scheme: we set $w_{ij} = 1$ if pixel $j$ is one of the four nearest neighbors of pixel$i$, and $w_{ij} = 0$ otherwise.\\ Both $\kappa$ maps contain $N=625$ pixels. Statistical significance was assessed using permutation testing with 5,999 random reassignments of pixel values to spatial locations for both $\kappa$ maps of different beam radii. In both cases the p-values were $< 10^{-4}$, indicating highly significant positive spatial autocorrelation.\clearpage


\bibliographystyle{MSP}
\bibliography{References}


\end{document}